\documentclass{aa}
\usepackage{txfonts}
\usepackage{graphicx}
\usepackage{rotating}

\begin{document}

\hyphenation{Fe-bru-ary Gra-na-da mo-le-cu-le mo-le-cu-les}


\title{Spectroscopic classification of red high proper motion objects 
in the Southern Sky 
\thanks{Based on observations collected with the VLT/FORS1 and ISAAC, 
the ESO\,3.6-m/EFOSC2 and the NTT/SOFI at the European Southern Observatory,
Paranal and La Silla, Chile (ESO programmes 63.L-0634, 65.L-0689, 68.C-0664, 
and 70.C-0568).}} 

\author{N. Lodieu \inst{1,2} \and
       R.-D. Scholz \inst{1} \and
       M. J. McCaughrean \inst{1,3} \and
       R. Ibata \inst{4} \and
       M. Irwin \inst{5} \and
       H. Zinnecker \inst{1}
}
\institute{Astrophysikalisches Institut Potsdam, An der Sternwarte 16, 
           14482 Potsdam, Germany 
           \and
           Department of Physics \& Astronomy, University of Leicester,
           University Road, Leicester LE1 7RH, UK
           \and 
           School of Physics, University of Exeter, Stocker Road,
           Exeter EX4 4QL, UK
           \and
	   Observatoire de Strasbourg, 11, rue de l'Universit\'e, 
           67000 Strasbourg, France 
           \and
	   Institute of Astronomy, Madingley Road, 
           Cambridge CB3 0HA, UK
}

\offprints{Nicolas Lodieu, nl41@star.le.ac.uk}

\date{Received \today / Accepted }

\titlerunning{Spectroscopic classification of red high proper motion objects}
\authorrunning{Lodieu et al.}

\abstract{
We present the results of spectroscopic follow-up observations for a sample 
of 71 red objects with high proper motions in the range 0.08--1.14 arcsec/yr 
as detected using APM and SSS measurements of multi-epoch photographic Schmidt 
plates. Red objects were selected by combining the photographic $B_JRI$ 
magnitudes with 2MASS near-infrared $JHK_{s}$ magnitudes. Some 50 of the 71
spectroscopically classified objects turn out to be late-type ($>$M6) 
dwarfs and in more detail, the sample includes 35 ultracool dwarfs with
spectral types between M8 and L2, some previously reported, as well as 
five M-type subdwarfs, including a cool esdM6 object, SSSPM J0500$-$5406. 
Distance estimates based on the spectral types and 2MASS $J$ magnitudes place 
almost all of the late-type ($>$M6) dwarfs within 50\,pc, with 25 objects
located inside the 25\,pc limit of the catalogue of nearby stars. Most of 
the early-type M dwarfs are located at larger distances of 100--200\,pc, 
suggesting halo kinematics for some of them. All objects with H$\alpha$ 
equivalent widths larger than 10\,\AA{} have relatively small tangential 
velocities ($<$\,50\,km/s). Finally, some late-type but blue objects 
are candidate binaries.
}
\maketitle


%
%
\section{Introduction}
\label{intro}
Although red dwarf stars of spectral type `M' account for more than 70\% 
of the stellar number density in the Solar neighbourhood and are estimated 
to contribute nearly half of the total stellar mass in the Galaxy (Henry et 
al.\ \cite{henry97}), they are not easily detectable in magnitude-limited sky 
surveys. Indeed, not a single known M dwarf, including the nearest star, 
Proxima Centauri (M5, d\,=\,1.3\,pc, $V$\,=\,11.1 mag), or the nearest M0 
dwarf, AX Microscopium (d\,=\,3.9\,pc, $V$\,=\,6.7 mag), is visible to the 
naked eye. Wide-field, deep 
images taken  with Schmidt telescopes preferentially reveal distant main 
sequence stars of earlier spectral types and more distant giants, but 
relatively few M dwarfs per field. One way of identifying these nearby, cool,
very low luminosity dwarf stars and their substellar counterparts is to 
compare multi-wavelength sky survey data and select on their characteristic 
colours. The drawback, however, is the presence of Galactic halo and compact
extragalactic sources with similar colours.

Proper motion surveys are an effective way to discriminate between distant 
giants and nearby dwarfs with similar colours. In general, large proper 
motions, e.g.\ $>$0.18 arcsec/yr, the lower limit in the New Luyten Two 
Tenths (NLTT) catalogue of $\sim$60,000 stars (Luyten \cite{luyten7980}),
indicate either small distances (d\,$<$\,50\,pc for members of the Galactic
disk population with typical tangential velocities $v_t$ of about 40\,km/s) 
and/or intrinsic high velocities (d\,$<$\,180\,pc for Galactic halo members 
with typical $v_{t}$\,$\sim$\,150\,km/s) (Scholz et al.\ \cite{scholz00}).
As a consequence, proper motion samples are highly biased 
towards halo members: according to Digby et al.\ (\cite{digby03}), the ratio 
of disk to halo stars can be as low as 4:1 for a proper motion-limited sample, 
compared to 400:1 for a true Solar neighbourhood volume-limited sample.
On the other hand, it is easier to rule out distant (halo) red giants in a 
sample of faint (e.g.\ $R$\,$>$\,13), red objects with proper motions 
exceeding the NLTT limit, since the proper motion would imply tangential 
velocities larger than $v_t$\,$>$\,10000\,km/s at distances of 
$d$\,$>$\,12\,kpc, a factor of 20 larger than the Galactic escape speed 
(cf.\ Leonard \& Tremaine \cite{leonard90}; Meillon et al.\ \cite{meillon97}). 

Despite the advantages of the high proper motion approach, the 3-D picture of 
the Solar neighbourhood is remarkably incomplete for low luminosity dwarfs 
due to the limiting magnitude and incompleteness of existing high proper 
motion catalogues, particularly in the southern sky and in the Galactic 
plane region. Within the 25\,pc horizon of the catalogue of nearby stars 
(Gliese \& Jahrei{\ss} \cite{gliese91}), possibly as many as 63\% of stellar 
systems remain undiscovered (Henry et al.\ \cite{henry02}), while the number 
of missing systems within 10\,pc is likely to be more than 30\% according to
Henry et al.\ (\cite{henry97}). Their prediction for the 10\,pc sample of 
$\sim$130 missing systems compared to the 229 known ones is based on the 
assumptions (1) that the density of stellar systems within 5\,pc (0.084 
systems pc$^{-3}$) extends to 25\,pc and (2) that the distribution of systems 
is isotropic. By comparing their 8\,pc sample to the 10\,pc sample, Reid 
et al.\ (\cite{reid03}) estimated an incompleteness in the latter of about 
25\%, with most of the missing stars are expected among the late-type M 
dwarfs. Even the 5\,pc sample cannot yet be considered complete with respect 
to these ultracool M dwarfs, as was recently demonstrated by the discovery of 
an M9 dwarf at about 4\,pc (Delfosse et al.\ \cite{delfosse01}) and an M6.5 
dwarf at about 3\,pc (Teegarden et al.\ \cite{teegarden03}).

Several groups have contributed to the completion of the nearby sample in
recent years. Since the identification of the twentieth nearest star, 
GJ\,1061 (Henry et al.\ \cite{henry97}), about 20 new M dwarfs have been 
discovered within 10\,pc. These include mid-M (M3.5 to M6.5) dwarfs (Scholz, 
Meusinger \& Jahrei{\ss} \cite{scholz01}; Reid \& Cruz \cite{reid02a}; Reid, 
Kilkenny, \& Cruz \cite{reid02b}; Reyl\'e et al.\ \cite{reyle02}; Henry et al.\
\cite{henry02}; Phan-Bao et al.\ \cite{phanbao03}; Teegarden et al.\ 
\cite{teegarden03}) and late-type (M7 to M9.5) dwarfs (Gizis et al.\ 
\cite{gizis00}; Delfosse et al.\ \cite{delfosse01}; McCaughrean, Scholz, 
\& Lodieu \cite{mjm02}; Reid et al.\ \cite{reid03}; L\'epine, Rich, \& 
Shara \cite{lepine03c}). Many of the mid-M dwarfs and some of the late-M 
dwarfs had already been catalogued as proper motion stars by Luyten (the 
Luyten Half Second [LHS] catalogue for stars with proper motions exceeding 
0.5 arcsec/yr; Luyten \cite{luyten79}), but their distances had not been
established. Initial distance estimates are generally arrived at via 
spectroscopic or photometric measurements, although these can be misleading
if the sources are multiples. Ultimately, confirmation via trigonometric 
parallax measurements is required.

Thanks to new near-infrared sky surveys, including the Two-Micron All Sky 
Survey (2MASS; Skrutskie et al.\ \cite{skrutskie97}) and the DEep 
Near-Infrared Survey (DENIS; Epchtein et al.\ \cite{epchtein97}), and deep
optical CCD surveys, including the Sloan Digital Sky Survey (SDSS; York et 
al.\ \cite{york00}), large numbers of objects even cooler than the latest-type 
M dwarfs have been recently discovered. These have led to the definition of 
two new spectral types, L and T, with accurate optical and near-infrared
classification schemes (Kirkpatrick et al.\ \cite{kirkpatrick99};
Mart\'{\i}n et al.\ \cite{martin99}; Burgasser et al.\ \cite{burgasser02};
Geballe et al.\ \cite{geballe02}). Many of these sources are also nearby
with high proper motions including, for example, the T dwarf binary
$\varepsilon$\,Indi\,Ba,Bb at 3.6\,pc, which has a proper motion of 4.8
arcsec/yr (Scholz et al.\ \cite{scholz03}; McCaughrean et al.\ 
\cite{mjm04}).

In this paper, we present the results of various photometric and spectroscopic 
follow-up observations of very red, high proper motion stars in the southern 
sky, selected as ultracool dwarf candidates during different stages of our 
ongoing high proper motion survey. The paper is organised as follows: 
\S\ref{sample} describes the various sample selection methods applied to 
reveal low-mass neighbours to the Sun. \S\ref{obs} outlines the optical and 
near-infrared photometry and spectroscopy observations made with several 
ESO telescopes, including the Very Large Telescope (VLT), New Technology 
Telescope (NTT), and 3.6-m. \S\ref{sp_class} explains the procedures employed 
in classifying our targets based on their optical and near-infrared properties.
A subsample of low-metallicity subdwarfs found in the course of the proper 
motion survey is presented in \S\ref{subdwarfs}. We discuss distance estimates 
in \S\ref{dist_est}, as well as the kinematics and measures of activity in
\S\ref{kinact}, for all objects detected in our search. Individual objects
with specific properties are discussed in more detail in \S\ref{special_obj},
including L dwarfs previously published by Lodieu, Scholz, \& McCaughrean
(\cite{lodieu02}) and Scholz \& Meusinger (\cite{scholz02b}), and M dwarfs 
reported by McCaughrean, Scholz, \& Lodieu (\cite{mjm02}).

%
\section{Selection of red proper motion stars}
\label{sample}
Detecting proper motion stars requires observational data spanning an
appropriately long period and the original Palomar Sky Survey (POSS-1)
led to the discovery of many proper motion sources when it was compared
with newer plate material. However, the relatively poor sensitivity of the 
POSS-1 plates meant that they were unable to pick up the very red and 
relatively faint L and T dwarfs: to date, none of the NLTT stars has been found 
to be an L dwarf or later. However, the more modern POSS-2, UK Schmidt, and
ESO Schmidt surveys provided substantial improvements in sensitivity at the
red end over several epochs, making it possible to discover nearby, high
proper motion L and T dwarfs via photographic material alone. This was 
successfully demonstrated by the discovery of the first field L dwarf, 
Kelu~1 (Ruiz et al.\ \cite{ruiz97}), other bright L dwarfs including
SSSPM\,J0829-1309 (Scholz \& Meusinger \cite{scholz02b}) and LSR\,0602+3910 
(Salim et al.\ \cite{salim03}), and the very nearby T dwarfs 
$\varepsilon$\,Indi\,Ba,Bb (Scholz et al.\ \cite{scholz03}).
Finally, comparison of data from digital optical and infrared 
sky surveys such as SDSS, 2MASS, and DENIS, with earlier plate material 
and/or second epoch digital data has also led to the discovery of large 
numbers of L and T dwarfs (Kirkpatrick et al.\ \cite{kirkpatrick99,kirkpatrick00};
Reid et al.\ \cite{reid00}; Scholz \& Meusinger 
\cite{scholz02b}; Tinney et al.\ \cite{tinney03}; Burgasser et al.\ 
\cite{burgasser03b,burgasser03c,burgasser04a}). All the nearest late-L and 
T~dwarfs have also been confirmed as large proper motion objects (Dahn 
et al.\ \cite{dahn02}; Vrba et al.\ \cite{vrba04}).

For the present work, we chose to use archival multi-epoch optical photographic 
plate material to search for new high proper motion sources, in common with 
some other groups. Several searches have been made in the poorly investigated
southern sky (e.g., Wroblewski \& Torres \cite{wroblewski94}; Wroblewski 
\& Costa \cite{wroblewski99}; Scholz et al.\ \cite{scholz00}; Ruiz et al.\
\cite{ruiz01}; Reyl\'e et al.\ \cite{reyle02}; Pokorny et al.\ 
\cite{pokorny03}; Hambly et al.\ \cite{hambly04}), while new, thorough 
surveys have also been made in the northern sky (L\'epine et al.\ 
\cite{lepine02}) and in the Galactic plane (L\'epine et al.\ \cite{lepine03d}).

%
%
\begin{figure*}[htb]
\begin{center}
\includegraphics[width=12.5cm, angle=270]{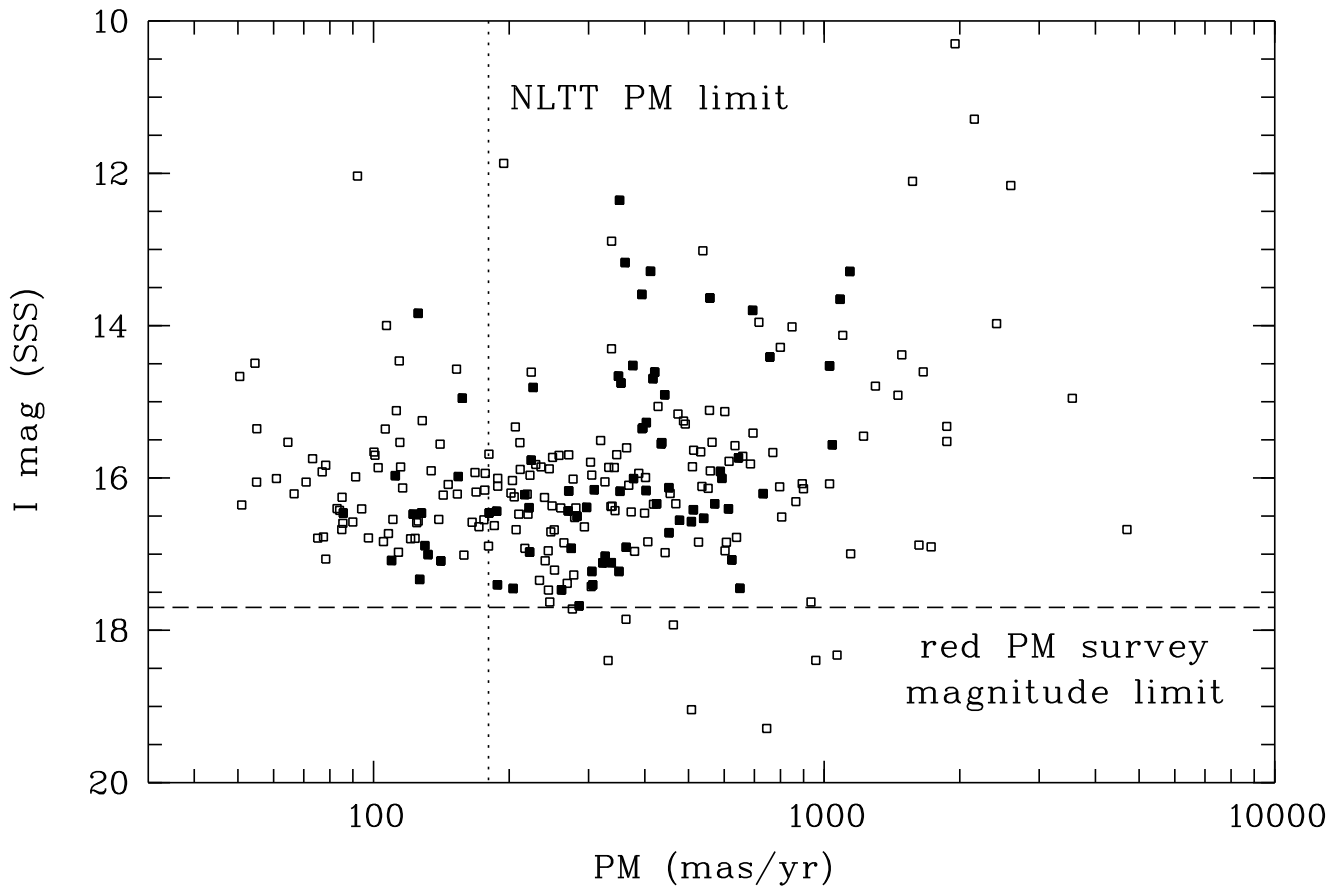}
\caption{
$I$-band magnitudes and proper motions for our full sample of about 270 red 
proper motion objects: the filled symbols represent the objects described in 
the current paper (Table \ref{tablecolours}). The handful of objects shown
below the red proper motion survey magnitude limit (dashed line) were 
originally discovered and classified as L and T dwarfs in deep digital 
surveys (SDSS, 2MASS, DENIS) based on their colours, but were then
recovered a posteriori in the SSS photographic data.
}
\label{Ipm}
\end{center}
\end{figure*}

Here, we started by selecting red candidates from the high proper motion 
survey initiated by Scholz et al.\ (\cite{scholz00}) to uncover missing 
nearby low-mass stars and brown dwarfs. This survey was originally based on 
APM (Automatic Plate Measuring) machine scans of UKST plates in two passbands 
($B_J$ and $R$) covering a few thousand square degrees (see, e.g., Reyl\'e 
et al.\ \cite{reyle02}): high proper motion sources discovered in this work
are given the prefix APMPM\@. More recently, the survey was extended to use
the multi-epoch astrometry and multi-colour ($B_J$, $R$, $I$) photographic 
photometry data in the first release of the SuperCOSMOS Sky Surveys        
(SSS; Hambly et al.\ 
\cite{hambly01a,hambly01b,hambly01c}): sources found this way are prefixed
SSSPM and the present sample derived from our recently completed southern
sky SSS survey is shown in Fig.\ \ref{Ipm}. It should be noted in passing
that the same SSS data are being used completely independently for the same 
purpose by other groups (Pokorny et al.\ \cite{pokorny03}; Hambly et al.\ 
\cite{hambly04}).


%
\begin{figure}[htb]
\begin{center}
\includegraphics[width=9.2cm, angle=0]{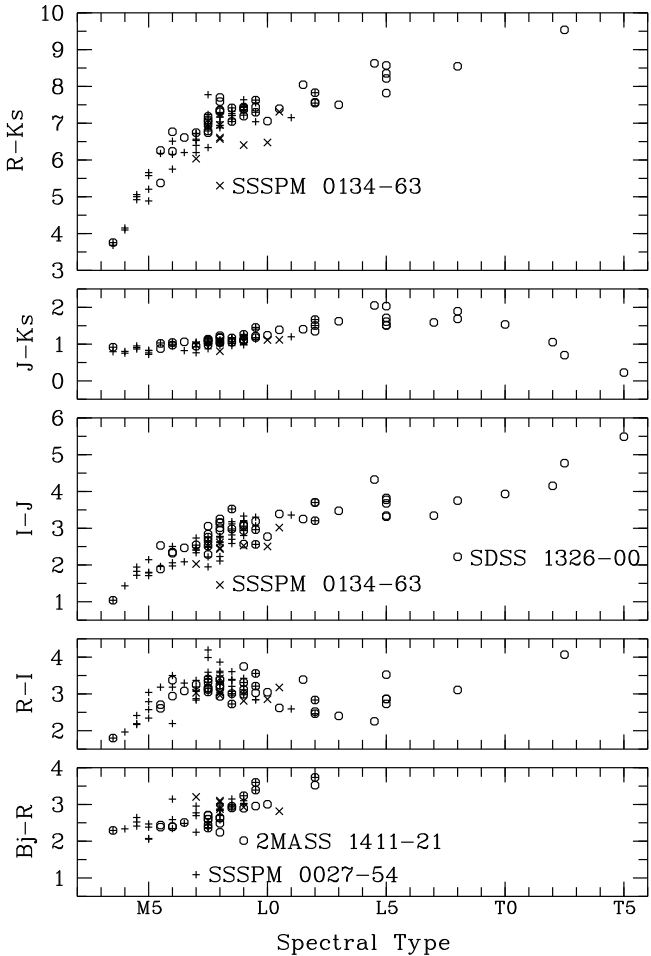}
\caption{
Colours of late M, L and T dwarfs as derived from SSS optical photometry
($B_{J}$, $R$, $I$) and 2MASS near-infrared photometry ($J$, $K_{s}$). Only 
objects detected in SSS data with spectral types obtained in this study 
($+$ optical, $\times$ near-infrared) and from the literature (open circles) 
are shown. Outliers with atypically blue colours for their spectral types 
are labelled by name (see \S\ref{coloutl}).
}
\label{colSpT}
\end{center}
\end{figure}

Our strategies for selecting nearby, ultracool dwarfs from the photographic
plate catalogues are constantly evolving and being refined, employing different 
combinations of proper motions, colours, magnitudes, and so on, as new 
photometric data become available and as our experience with spectroscopically 
confirmed sources is folded back into the procedure. The main search 
strategies employed to date can be summarised as follows (in chronological 
order):
\begin{enumerate}
\item 
APM data were available for only one pair of plates ($B_{J}$ and $R$) per 
field, with a typical epoch difference between 5 and 15 years. We selected
objects shifted by more than 5 arcsec (corresponding to a lower proper
motion limit of 0.33 arcsec/yr; for more details see Scholz et al.\ 
\cite{scholz00}) and with $B_J - R > 2.5$. Since the photographic $B_J - R$ 
colour is only weakly correlated with spectral type (see Fig.\ \ref{colSpT}, 
cf.\ Fig.\ 3 in Kirkpatrick, Henry \& Irwin \cite{kirkpatrick97}), it was
hard to predict the optical and near-infrared fluxes for each source and 
thus we obtained optical ($R$, $I$) and near-infrared ($J$, $H$, $K_s$) 
photometry from VLT pre-imaging in order to ensure accurate exposure times 
for the subsequent spectroscopy (see \S\ref{vltobs}). 
\item 
The initial 1999 release of SSS data included about 5000 square degrees at the 
South Galactic Cap ($b < -60^{\circ}$) with $B_J$ and $R$ photometry from the 
UKST plates as well as additional epoch measurements from UKST $I$ and ESO 
Schmidt $R$ plates. The proper motions given in the SSS catalogue were 
obtained from UKST $B_J$ and $R$ plates only, for all objects matched between 
the two passbands within a search radius of 3 arcsec. In order to find sources
with larger proper motions, we selected unmatched objects which were picked up
along a straight line on at least three plates with consecutive epochs (for 
more details see Lodieu et al.\ \cite{lodieu02}). The proper motions were then 
refined by including all available positional information. Ultracool dwarf
candidates were identified from these confirmed high proper motion objects 
by looking for the largest $B_J - R$ and $R - I$ indices.
\item
The SSS UKST $B_J$ and $R$ all-southern-sky surveys were completed in 2001, 
with additional UKST $I$ survey and ESO $R$ survey data added subsequently.
The catalogue photometric zero points were re-derived, resulting in significant
changes in the $B_J - R$ and $R - I$ colours. Our revised search strategy 
then concentrated on the $R$ and $I$ plates alone, neglecting the blue $B_J$ 
plates which are much less sensitive to extremely red objects.
\item
As an alternative strategy, we also looked for very red candidates among 
objects originally matched in all three passbands within a search radius of 
6 arcsec and having a proper motion of $>$100\,mas/yr in the SSS database
(for more details see Scholz \& Meusinger \cite{scholz02b}). We subsequently
reduced this limit to 50\,mas/yr when the SSS proper motion determinations 
incorporated all plate measurements in all passbands, i.e., they were no 
longer based on the UKST $B_J$ and $R$ scans alone.
\item
The SSS $I$ band survey was completed at the end of 2002. As there is an 
overlap of 0.5 degrees on each side of an UKST plate with its neighbours,
roughly one third of the southern sky SSS $I$ band measurements had at least
two different epochs. The epoch difference is non-uniform and generally only
a few years, but nevertheless, these overlapping areas can be used to search 
for very high proper motion nearby brown dwarfs which may not show up on the 
$B_J$ and $R$ plates due to their extremely red colours. This new search method 
led to the discovery of nearest known brown dwarf, $\varepsilon$\,Indi\,B 
(Scholz et al.\ \cite{scholz03}), subsequently resolved into a binary
with spectral types T1 and T6 (McCaughrean et al.\ \cite{mjm04}).
\end{enumerate}

\vspace{-0.1cm}

For each of the five search strategies outlined above, we applied a magnitude 
limit at the faint end that ensured a sample of no more than a few thousand 
candidates. Source-by-source visual inspection then reduced the final sample
to a few hundred objects for follow-up spectroscopic observations. 
Fig.\  \ref{Ipm} shows the $I$ versus proper motion distribution for our
current full sample of about 270 red proper motion objects, where filled 
symbols represent the objects described in the present paper. The effective
faint magnitude limit of the combined sample of red proper motion objects is 
$I\sim 17.7$ mag. Sources in Fig.\ \ref{Ipm} below this limit are ultracool 
L and T dwarfs culled from the literature which we subsequently found to be
present in the SSS data but which were not picked up by our search strategies,
mainly because they were below our applied magnitude limits.

Fig.\ \ref{Ipm} also illustrates that our search strategy is sensitive to
sources faint enough to be low-mass dwarfs even down to a lower proper motion 
limit of $\sim$50\,mas/yr: candidates with proper motions this small were selected
only if they also exhibited extremely red colours typical of late-M and early-L
dwarfs in order to exclude more distant higher-mass sources. Our selected
proper motion sample is also generally less complete at the lowest proper 
motions ($<$300~mas/yr) due to the constraints set in the first search 
strategy. The sub-sample described in the present
paper has a lower limit of about 80\,mas/yr. At the other extreme, objects 
with proper motions larger than $\sim$\,1 arcsec/yr were included as 
interesting candidates even if they had only moderately red colours, as these 
may be nearby mid-M dwarfs. The objects with the largest proper motions in
Fig.\ \ref{Ipm}, above the 1.14 arcsec/yr upper limit in the present paper, 
were all newly discovered or recovered within the framework of our high proper 
motion survey. Some of these have already been published separately (Scholz 
et al.\ \cite{scholz00,scholz02a,scholz03,scholz04b,scholz04c}), whereas others are 
targets for upcoming spectroscopic follow-up. Five were previously discovered 
and classified by other surveys (Goldman et al.\ \cite{goldman99}; Mart\'{\i}n 
et al.\ \cite{martin99}; Delfosse et al.\ \cite{delfosse01}; L\'epine et al.\
\cite{lepine03b}; Hambly et al.\ \cite{hambly04}).

Fig.\ \ref{colSpT} shows photographic optical and 2MASS near-infrared colours
as a function of spectral type for all objects spectroscopically classified 
in the present study, together with all M, L, and T dwarfs from the literature 
which could be identified in the SSS data. These detections of previously 
known objects include three further T dwarfs, SDSS 0423$-$04 (T0; Geballe 
et al.\ \cite{geballe02}), SDSS 1254$-$01 (T2; Leggett et al.\ 
\cite{leggett00}), and 2MASS 0559$-$14 (T5; Burgasser et al.\ 
\cite{burgasser00}), in addition to $\varepsilon$\,Indi\,B and ten L dwarfs 
already described by Scholz \& Meusinger (\cite{scholz02b}).

The optical--infrared colour indices of the M, L, and T dwarfs measured by
the SSS (Fig.\ \ref{colSpT}) increase monotonically with spectral types. Pure 
optical $R - I$ and infrared $J - K_s$ colours saturate at about M8 and L5,
respectively, but then turn bluer with later spectral type: the $R - I$
colour then starts becoming redder again at about L5. The smallest colour 
changes appear in $J - K_s$, where a high photometric accuracy yields a 
small dispersion. Conversely, the largest effect is seen in $R - K_s$, 
although its use as a selector would lead to the exclusion of some L and 
T dwarfs not visible shortward of the $I$ band. Finally, there are some 
clear outliers in the optical and optical--infrared colours labelled by 
name in Fig.\ \ref{colSpT}: these objects will be discussed in \S\ref{coloutl}.

The sample described in this paper contains a subset of 79 red high proper
motion objects selected from our full candidate list (Table~\ref{tablecolours}).
As described in the following sections, we have obtained spectra with the VLT,
NTT, and ESO 3.6-m at optical wavelengths for 58 of them and in the 
near-infrared for 27, the latter including 19 objects already observed in 
the optical. VLT optical and near-infrared photometry was obtained for a 
subsample of objects as catalogued here, in addition to the $B_J$, $R$, and 
$I$ photographic plate measurements. Some comparison stars with known spectral 
types were also observed with identical setups: LHS\,517 (M3.5), 
2MASSW\,J0952$-$19 (M7.0), LP647-13 (M7.5), BRI\,B0021$-$0214 (M9.5), LP944-20 (M9.5),
and Kelu\,1 (L2.0). In addition, some red NLTT stars lacking spectral types were 
also observed. Finally, some of the objects in the sample were subsequently 
identified with previously known proper motion sources, including ESO\,207-61, 
LHS\,2555a, CE\,303, LP\,859-1, CE\,352, [HB88]\,M18, and [HB88]\,M12. 

%
%
\section{Observations and data reduction}
\label{obs}
\subsection{VLT service mode observations}
\label{vltobs}
An initial subsample of very red proper motion objects were observed
photometrically and spectroscopically on the ESO VLT UT1 in service mode
in 1999 (P63.L-0634) and in 2000 (P65.L-0689) in the optical with the Focal
Reducer/Low Dispersion Spectrograph FORS1 and in the near-infrared with the
Infrared Spectrometer And Array Camera (ISAAC). The observations were
conducted in dark and grey time in the optical and near-infrared,
respectively, and under good seeing conditions ($\leq 0.8$ arcsec) as
requested.
\subsubsection{VLT/FORS1 imaging}
\label{FORS1_Imaging}
FORS1 employs a 2048$\times$2048 pixel thinned TK CCD detector: the image
scale of 0.20 arcsec yields a field-of-view of 6.8$\times$6.8 arcmin.
A series of three dithered positions was exposed for 10 and 5 seconds
each in each of the Bessell $R$ and $I$ filters, respectively, in order
to derive more accurate magnitudes and colours than were available from
the plate material. Standard stars were observed each night in order to
calibrate the photometry. Data reduction was standard, with each frame
bias-subtracted and flat-fielded prior to aperture photometry.
Table \ref{table_img_OBs} lists the mean magnitude calculated from 
three independent measurements, the exposure times, and the specific
observing dates for 32 sources in the initial subsample.

%
%
\begin{table*}
\thispagestyle{headings}
\footnotesize
\begin{center}
\caption[Names, coordinates, proper motions, and magnitudes of 
red proper motion objects]{\footnotesize
List of new, red proper motion objects selected as explained in the text. 
Columns\,1 and 2: reference number and name of the object;
Columns\,3 and 4: equinox J2000 coordinates at the latest available epoch;
Column\,5: epoch of position;
Columns\,6--9: proper motion in mas/yr and associated errors;
Columns\,10--12: $B_{J}$, $R$, $I$ magnitudes from SSS; 
Columns\,13--15: $J$, $H$, $K_{s}$ magnitudes from 2MASS\@.
Some comparison stars with well-determined spectral types are included
in the list, namely LHS\,517 (M3.5), 2MASS\,J0952$-$19 (M7.0), LP647-13 (M9),
BRI\,B0021$-$0214 (M9.5), and Kelu\,1 (L2.0). Notes $^a$ and $^b$ indicate 
that the object is also mentioned in the proper motion catalogue of 
Pokorny, Jones, \& Hambly (\cite{pokorny03}) and in the B1.0 catalogue 
(Monet et al.\ 2003), respectively.
}
 \tiny
 \begin{tabular}{|@{\hspace{1mm}}rl|@{\hspace{3mm}}cc|c|@{\hspace{3mm}}rrrr|c@{\hspace{3mm}}c@{\hspace{3mm}}c@{\hspace{3mm}}c@{\hspace{3mm}}c@{\hspace{3mm}}c|}
 \hline
RN & Name    & \multicolumn{2}{c|}{Coordinates (J2000)} & Epoch & \multicolumn{4}{c|}{Proper Motions} & \multicolumn{6}{c|}{Optical and Infrared Magnitudes} \cr
 \hline
     &   & R.A. & Declination & & $\mu_{\alpha}\cos\delta$  & $\mu_{\delta}$ & $\sigma$x & $\sigma$y & $B_{J}$ & $R$ & $I$ & $J$ & $H$ & $K_s$ \cr
 \hline
01 & SSSPM J0006$-$2157     $^{a,b}$ &  0$:$05$:$48.46 & $-$21$:$57$:$19.7 & 1999.60 &  $+$721 &  $-$128 &  7 &  2 & 22.56 & 19.41 & 16.21 & 13.27 & 12.62 & 12.20 \cr 
02 &   BRI B0021$-$0214     $^{  b}$ &  0$:$24$:$24.61 & $-$01$:$58$:$19.5 & 2001.64 &   $-$93 &  $+$127 &  3 &  6 & 21.76 & 18.17 & 14.95 & 11.99 & 11.08 & 10.54 \cr 
03 & SSSPM J0027$-$5402     $^{  b}$ &  0$:$27$:$23.43 & $-$54$:$01$:$46.1 & 1999.80 &  $+$415 &   $+$41 & 18 &  4 & 19.15 & 18.06 & 14.70 & 12.36 & 11.72 & 11.34 \cr 
04 & SSSPM J0030$-$3427     $^{  a}$ &  0$:$30$:$10.23 & $-$34$:$26$:$55.5 & 2000.72 &  $-$102 &  $-$279 &  7 &  2 & 22.12 & 19.20 & 16.39 & 13.86 & 13.19 & 12.79 \cr 
05 &    LP 645-52         $^{  b}$ &  0$:$35$:$41.68   & $-$03$:$21$:$30.8 & 1998.71 &  $+$428 &   $-$84 &  3 &  1 & 20.26 & 17.88 & 15.54 & 13.75 & 13.28 & 12.99 \cr 
06 & APMPM J0057$-$7604     $^{  b}$ &  0$:$56$:$53.38 & $-$76$:$03$:$43.3 & 1998.60 &  $+$160 &  $-$389 &  4 & 13 & 19.88 & 16.91 & 14.61 & 12.57 & 11.99 & 11.62 \cr 
07 & SSSPM J0109$-$5101     $^{  b}$ &  1$:$09$:$01.50 & $-$51$:$00$:$49.4 & 1999.81 &  $+$209 &   $+$86 &  2 &  7 & 21.19 & 18.21 & 14.81 & 12.23 & 11.54 & 11.09 \cr 
08 & SSSPM J0109$-$4955     $^{  b}$ &  1$:$09$:$09.18 & $-$49$:$54$:$53.2 & 1999.81 &   $+$86 &  $+$128 &  9 &  8 & 22.32 & 19.40 & 15.98 & 13.55 & 12.88 & 12.45 \cr 
09 &    LP 647-13         $^{  b}$ &  1$:$09$:$51.17   & $-$03$:$43$:$26.4 & 1998.71 &  $+$354 &   $+$13 &  4 &  5 & 21.10 & 17.87 & 14.75 & 11.69 & 10.93 & 10.43 \cr 
10 & SSSPM J0124$-$4240     $^{a,b}$ &  1$:$23$:$59.05 & $-$42$:$40$:$07.3 & 2000.63 &  $-$145 &  $-$229 &  4 &  7 & 22.16 & 19.35 & 16.17 & 13.15 & 12.47 & 12.04 \cr 
11 & SSSPM J0125$-$6546     $^{   }$ &  1$:$24$:$49.63 & $-$65$:$46$:$33.6 & 1999.90 &   $+$98 &   $+$82 &  7 &  8 & 22.70 & 19.49 & 16.46 & 14.43 & 13.81 & 13.46 \cr 
12 & SSSPM J0134$-$6315     $^{   }$ &  1$:$33$:$32.44 & $-$63$:$14$:$41.8 & 1999.90 &   $+$77 &   $-$81 &  8 &  9 & 22.11 & 19.01 & 15.97 & 14.51 & 14.02 & 13.70 \cr 
13 &    LP 769-14         $^{a,b}$ &  1$:$59$:$17.41   & $-$17$:$30$:$08.7 & 2000.81 &  $-$119 &  $-$145 &  3 &  6 & 21.15 & 18.63 & 16.44 & 14.61 & 13.96 & 13.70 \cr 
14 & SSSPM J0204$-$3633     $^{a,b}$ &  2$:$04$:$22.13 & $-$36$:$32$:$30.8 & 2000.72 &  $+$216 &   $-$59 &  9 & 26 &        & 19.96 & 15.77 & 13.27 & 12.60 & 12.19 \cr 
15 & APMPM J0207$-$7214     $^{  b}$ &  2$:$07$:$05.28 & $-$72$:$14$:$06.9 & 1999.86 &  $+$282 &  $-$120 &  6 &  5 & 21.07 & 18.59 & 17.40 & 16.50 & 15.93 & 15.82 \cr 
16 & APMPM J0207$-$3722     $^{a,b}$ &  2$:$07$:$14.08 & $-$37$:$21$:$50.2 & 2000.72 &  $+$422 &  $+$134 &  2 &  6 & 20.79 & 18.10 & 14.91 & 12.44 & 11.83 & 11.38 \cr 
17 & SSSPM J0215$-$4804     $^{a,b}$ &  2$:$14$:$48.14 & $-$48$:$04$:$25.3 & 2000.01 &  $+$118 &  $-$332 &  9 &  5 & 22.20 & 19.14 & 16.17 & 13.56 & 12.96 & 12.52 \cr 
18 & SSSPM J0219$-$1939     $^{   }$ &  2$:$19$:$28.07 & $-$19$:$38$:$41.6 & 2000.89 &  $+$195 &  $-$174 &  4 &  5 &        & 20.06 & 17.47 & 14.11 & 13.34 & 12.91 \cr 
19 & SSSPM J0222$-$5412     $^{   }$ &  2$:$21$:$54.94 & $-$54$:$12$:$05.4 & 1999.82 &  $+$107 &   $-$14 &  7 &  8 &        & 20.30 & 17.08 & 13.90 & 13.22 & 12.66 \cr 
20 & SSSPM J0231$-$4122     $^{   }$ &  2$:$31$:$22.25 & $-$41$:$21$:$50.7 & 1999.66 &  $+$301 &  $-$127 &  4 & 28 &        & 20.10 & 17.03 & 13.85 & 13.27 & 12.89 \cr 
21 & APMPM J0232$-$4437     $^{  b}$ &  2$:$32$:$08.96 & $-$44$:$37$:$00.1 & 1998.92 &  $+$384 &  $+$284 &  5 &  5 & 21.42 & 19.00 & 16.56 & 14.45 & 13.82 & 13.74 \cr 
22 & APMPM J0244$-$5203     $^{a,b}$ &  2$:$44$:$00.15 & $-$52$:$02$:$43.0 & 1999.77 &  $+$305 &   $-$11 &  7 &  3 & 21.49 & 18.96 & 17.23 & 15.92 & 15.41 & 14.99 \cr 
23 & SSSPM J0306$-$3648     $^{a,b}$ &  3$:$06$:$11.59 & $-$36$:$47$:$52.8 & 2000.00 &  $-$180 &  $-$670 &  9 &  8 & 20.55 & 17.67 & 13.80 & 11.69 & 11.07 & 10.63 \cr 
24 & SSSPM J0327$-$4236     $^{  b}$ &  3$:$26$:$32.78 & $-$42$:$36$:$08.3 & 2000.77 &  $+$286 &   $-$22 & 25 & 28 &        & 20.53 & 16.92 & 14.22 & 13.61 & 13.21 \cr 
25 & APMPM J0331$-$2349     $^{a,b}$ &  3$:$30$:$39.10 & $-$23$:$48$:$45.9 & 1998.89 &  $+$621 &   $+$58 &  8 &  5 & 21.30 & 18.87 & 17.08 & 15.80 & 15.31 & 14.94 \cr 
26 &    LP 888-18         $^{a,b}$ &  3$:$31$:$30.25   & $-$30$:$42$:$38.8 & 1999.94 &   $+$40 &  $-$392 &  5 &  6 & 19.91 & 17.21 & 13.59 & 11.36 & 10.70 & 10.26 \cr 
27 &    LP 944-20         $^{  a}$ &  3$:$39$:$35.21   & $-$35$:$25$:$44.0 & 1998.93 &  $+$302 &  $+$280 &  6 & 10 & 20.24 & 16.84 & 13.29 & 10.73 & 10.02 &  9.55 \cr 
28 &    LP 775-31         $^{  b}$ &  4$:$35$:$16.12   & $-$16$:$06$:$57.4 & 1998.90 &  $+$156 &  $+$315 &  3 &  4 & 18.85 & 16.34 & 12.35 & 10.41 &  9.78 &  9.35 \cr 
29 &    LP 655-48         $^{  b}$ &  4$:$40$:$23.33   & $-$05$:$30$:$07.9 & 2001.79 &  $+$339 &  $+$126 &  2 &  2 & 18.85 & 16.50 & 13.17 & 10.66 &  9.99 &  9.55 \cr 
30 & SSSPM J0500$-$5406     $^{a,b}$ &  5$:$00$:$15.77 & $-$54$:$06$:$27.3 & 1999.84 &  $+$207 & $-$1022 &  9 &  2 & 20.10 & 17.31 & 15.56 & 14.44 & 14.12 & 13.97 \cr 
31 & SSSPM J0511$-$4606     $^{   }$ &  5$:$11$:$01.63 & $-$46$:$06$:$01.5 & 1999.77 &   $+$53 &  $+$121 &  9 &  6 &        & 19.97 & 17.01 & 13.89 & 13.19 & 12.71 \cr 
32 & APMPM J0536$-$5358     $^{  b}$ &  5$:$36$:$21.04 & $-$53$:$58$:$29.7 & 1999.88 &  $+$289 &  $-$243 &  5 &  4 & 21.64 & 19.02 & 16.01 & 13.93 & 13.30 & 12.90 \cr 
33 &   ESO 207-61         $^{   }$ &  7$:$07$:$53.27   & $-$49$:$00$:$50.3 & 2000.15 &   $-$34 &  $+$401 & 14 &  8 &        & 19.48 & 16.17 & 13.23 & 12.54 & 12.10 \cr 
34 & SSSPM J0829$-$1309     $^{  b}$ &  8$:$28$:$34.11 & $-$13$:$09$:$20.1 & 2001.28 &  $-$593 &   $+$14 &  6 &  7 & 22.58 & 18.84 & 16.01 & 12.80 & 11.85 & 11.30 \cr 
35 &    LP 314-67         $^{  b}$ &  9$:$48$:$05.16   & $+$26$:$24$:$18.9 & 1999.07 &  $-$143 &  $-$438 & 13 &  2 &        & 18.03 &        & 15.59 & 15.03 & 14.85 \cr 
36 & 2MASSW J0952$-$19$^{  b}$ &  9$:$52$:$21.88       & $-$19$:$24$:$31.9 & 1998.32 &   $-$76 &  $-$100 &  3 &  2 & 19.37 & 17.02 & 13.84 & 11.86 & 11.26 & 10.87 \cr 
37 &    LP 614-35         $^{  b}$ & 12$:$07$:$51.63   & $-$00$:$52$:$32.0 & 1999.07 &  $-$188 &   $-$11 &  2 &  4 & 20.92 & 18.35 & 17.40 & 16.15 & 15.50 & 15.30 \cr 
38 & APMPM J1212$-$2126     $^{  b}$ & 12$:$11$:$31.86 & $-$21$:$25$:$43.4 & 1998.29 &  $-$330 &   $-$68 &  8 &  8 & 21.07 & 18.59 & 17.11 & 16.27 & 15.71 & 15.60 \cr 
39 & APMPM J1222$-$2452     $^{  b}$ & 12$:$22$:$26.55 & $-$24$:$52$:$15.8 & 1998.50 &  $-$442 &   $+$96 & 12 &  4 & 21.17 & 18.70 & 16.13 & 14.33 & 13.85 & 13.49 \cr 
40 &   LHS 2555a          $^{  b}$ & 12$:$24$:$46.13   & $-$32$:$00$:$16.8 & 1999.26 &  $-$598 &  $-$243 & 12 &  9 & 19.54 & 17.29 & 15.73 & 14.31 & 13.79 & 13.57 \cr 
41 & APMPM J1251$-$2121     $^{  b}$ & 12$:$50$:$52.65 & $-$21$:$21$:$13.6 & 2000.20 &  $+$435 &  $-$349 &  4 &  3 & 19.40 & 17.03 & 13.64 & 11.16 & 10.55 & 10.13 \cr 
42 &  Kelu 1              $^{  b}$ & 13$:$05$:$40.19   & $-$25$:$41$:$05.9 & 1998.41 &  $-$322 &   $-$19 & 10 &  5 &        & 19.58 & 17.11 & 13.41 & 12.39 & 11.75 \cr 
43 &    CE 303            $^{  b}$ & 13$:$09$:$21.85   & $-$23$:$30$:$35.0 & 1998.32 &    $+$8 &  $-$376 &  9 &  2 & 20.41 & 17.83 & 14.52 & 11.78 & 11.08 & 10.67 \cr 
44 &    CE 352            $^{  b}$ & 13$:$40$:$38.77   & $-$30$:$32$:$02.7 & 2000.23 &  $-$335 &  $-$103 &  7 &  4 & 20.58 & 18.26 & 17.23 & 15.70 & 15.27 & 15.04 \cr 
45 &    LP 859-1          $^{  b}$ & 15$:$04$:$16.21   & $-$23$:$55$:$56.4 & 1998.48 &  $-$339 &   $-$85 & 14 &  3 & 20.26 & 17.82 & 14.66 & 12.01 & 11.38 & 11.03 \cr 
46 &   LHS 3141B          $^{  b}$ & 15$:$59$:$37.99   & $-$22$:$26$:$12.8 & 1999.32 &  $+$193 &  $-$538 &  8 & 21 & 21.65 & 19.41 & 16.34 & 13.61 & 13.12 & 12.84 \cr 
47 & SSSPM J1926$-$4311     $^{   }$ & 19$:$26$:$08.59 & $-$43$:$10$:$56.3 & 1999.53 &  $-$167 & $-$1072 & 15 & 20 & 18.75 & 16.69 & 13.65 & 11.94 & 11.42 & 11.12 \cr 
48 & APMPM J1957$-$4216     $^{  b}$ & 19$:$56$:$57.61 & $-$42$:$16$:$23.5 & 2000.57 &  $+$149 & $-$1017 &  4 &  6 & 19.39 & 17.32 & 14.53 & 12.38 & 11.99 & 11.66 \cr 
49 &    LP 815-21         $^{  b}$ & 20$:$28$:$04.52   & $-$18$:$18$:$57.5 & 1998.44 &  $-$105 &  $-$175 &  5 &  6 & 20.38 & 18.11 & 17.45 & 16.10 & 15.91 & 15.28 \cr 
50 & SSSPM J2033$-$6919     $^{  b}$ & 20$:$32$:$32.91 & $-$69$:$18$:$59.1 & 2000.43 &  $+$228 &  $-$459 & 10 &  5 & 22.59 & 19.75 & 16.42 & 13.64 & 12.98 & 12.58 \cr 
51 & APMPM J2036$-$4936     $^{  b}$ & 20$:$35$:$49.96 & $-$49$:$36$:$07.7 & 1999.71 &   $-$86 &  $-$416 &  4 &  8 & 21.17 & 18.75 & 16.34 & 14.62 & 14.16 & 13.75 \cr 
52 & SSSPM J2052$-$4759     $^{  b}$ & 20$:$52$:$28.08 & $-$47$:$58$:$44.2 & 1999.78 &    $-$7 &  $-$435 &  5 &  7 & 21.67 & 18.82 & 15.55 & 12.94 & 12.29 & 11.88 \cr 
53 & SSSPM J2059$-$8018     $^{   }$ & 20$:$59$:$02.19 & $-$80$:$17$:$36.9 & 2000.66 &  $+$361 &   $-$46 &  6 & 29 &        & 20.50 & 16.91 & 14.28 & 13.69 & 13.41 \cr 
54 & SSSPM J2101$-$5110     $^{   }$ & 21$:$01$:$29.49 & $-$51$:$10$:$02.9 & 1999.64 &   $+$69 &  $-$123 &  7 & 16 &        & 20.27 & 17.09 & 15.13 & 14.41 & 14.09 \cr 
55 & [HB88] M18            $^{  b}$ & 21$:$18$:$31.74  & $-$45$:$05$:$52.2 & 1999.71 &  $+$388 &  $-$475 & 14 & 15 & 22.36 & 19.41 & 16.41 & 13.43 & 12.77 & 12.37 \cr 
56 & [HB88] M12            $^{  b}$ & 21$:$31$:$14.14  & $-$42$:$24$:$14.3 & 1999.63 &   $+$48 &   $-$71 &  6 &  9 & 21.27 & 18.63 & 16.46 & 14.52 & 13.92 & 13.57 \cr 
57 &    LP 819-9          $^{  b}$ & 21$:$59$:$30.91   & $-$15$:$54$:$16.5 & 2000.78 &    $+$0 &  $-$283 &  2 &  4 & 20.80 & 18.47 & 16.50 & 15.07 & 14.59 & 14.31 \cr 
58 &   LHS 517            $^{  b}$ & 22$:$09$:$40.29   & $-$04$:$38$:$26.7 & 1999.39 & $+$1125 &   $-$24 &  2 &  1 & 11.64 &  9.35 &  7.55 &  6.51 &  5.90 &  5.59 \cr 
59 & SSSPM J2229$-$6931     $^{  b}$ & 22$:$29$:$23.65 & $-$69$:$30$:$56.9 & 2000.58 &   $+$47 &  $-$217 &  9 & 10 &        & 19.83 & 16.97 & 14.47 & 13.76 & 13.35 \cr 
60 &LDS4980 B              $^{  b}$ & 22$:$35$:$58.17  & $+$07$:$57$:$13.9 & 2000.59 &  $-$112 &  $-$208 &  2 &  6 &        & 18.35 &        & 15.05 & 14.53 & 14.25 \cr 
61 &LDS4980 A              $^{  b}$ & 22$:$36$:$00.63  & $+$07$:$56$:$03.5 & 2000.59 &  $-$103 &  $-$210 &  9 &  2 &        & 17.54 &        & 14.65 & 14.26 & 13.86 \cr 
62 & SSSPM J2240$-$4253     $^{  b}$ & 22$:$40$:$26.97 & $-$42$:$53$:$18.4 & 2000.73 &   $-$35 &  $-$539 & 27 &  9 &        & 19.67 & 16.53 & 13.76 & 13.19 & 12.80 \cr 
63 & SSSPM J2257$-$5208     $^{   }$ & 22$:$57$:$31.68 & $-$52$:$08$:$26.3 & 1999.82 &   $+$57 &  $+$113 & 11 & 16 &        & 20.20 & 17.33 & 14.93 & 14.37 & 14.00 \cr 
64 & SSSPM J2258$-$4639     $^{   }$ & 22$:$57$:$49.26 & $-$46$:$38$:$44.5 & 2000.73 &  $-$185 &  $+$197 & 11 &  4 &        & 19.81 & 16.43 & 13.61 & 12.93 & 12.60 \cr 
65 & SSSPM J2307$-$5009     $^{  b}$ & 23$:$06$:$58.76 & $-$50$:$08$:$58.9 & 1999.84 &  $+$452 &   $+$25 & 10 &  6 & 22.80 & 19.70 & 16.72 & 13.39 & 12.70 & 12.24 \cr 
66 & SSSPM J2310$-$1759     $^{   }$ & 23$:$10$:$18.46 & $-$17$:$59$:$09.0 & 1998.50 &   $-$23 &  $-$271 & 36 & 24 &        & 20.52 & 17.68 & 14.38 & 13.58 & 12.97 \cr 
67 & SSSPM J2319$-$4919     $^{  b}$ & 23$:$18$:$46.14 & $-$49$:$19$:$18.0 & 2000.50 &  $+$216 &   $-$16 &  9 &  3 & 22.12 & 19.25 & 16.22 & 13.76 & 13.07 & 12.68 \cr 
68 & SSSPM J2322$-$6358     $^{   }$ & 23$:$22$:$05.69 & $-$63$:$57$:$58.0 & 1999.88 &  $+$121 &   $-$19 & 11 &  5 &        & 19.54 & 16.48 & 14.26 & 13.65 & 13.20 \cr 
69 & APMPM J2330$-$4737     $^{   }$ & 23$:$30$:$16.12 & $-$47$:$36$:$45.9 & 2000.79 &  $-$578 &  $-$983 &  2 &  2 & 19.38 & 16.79 & 13.29 & 11.23 & 10.64 & 10.28 \cr 
70 & APMPM J2331$-$2750     $^{  b}$ & 23$:$31$:$21.74 & $-$27$:$49$:$50.0 & 1999.44 &   $+$85 &  $+$753 &  4 &  3 & 20.42 & 17.89 & 14.41 & 11.65 & 11.06 & 10.65 \cr 
71 & SSSPM J2335$-$6913     $^{  b}$ & 23$:$35$:$19.59 & $-$69$:$13$:$17.0 & 2000.76 &   $+$66 &  $-$168 & 10 &  3 & 22.25 & 19.29 & 16.46 & 13.92 & 13.25 & 12.90 \cr 
72 & APMPM J2344$-$2906     $^{   }$ & 23$:$43$:$31.98 & $-$29$:$06$:$27.1 & 1998.85 &  $+$331 &  $-$217 & 10 &  6 & 21.12 & 18.64 & 15.34 & 13.26 & 12.75 & 12.43 \cr 
73 & SSSPM J2345$-$6810     $^{  b}$ & 23$:$44$:$57.97 & $-$68$:$09$:$39.8 & 2000.77 &  $+$206 &   $-$81 & 11 &  4 & 22.26 & 19.49 & 16.39 & 13.98 & 13.36 & 12.96 \cr 
74 & APMPM J2347$-$3154     $^{  b}$ & 23$:$46$:$54.71 & $-$31$:$53$:$53.2 & 1998.95 &  $+$424 &  $-$408 &  3 &  4 & 22.51 & 19.49 & 15.91 & 13.28 & 12.68 & 12.20 \cr 
75 & SSSPM J2352$-$2538     $^{  a}$ & 23$:$51$:$50.44 & $-$25$:$37$:$36.6 & 1999.60 &  $+$354 &  $+$193 &  9 &  8 & 21.73 & 18.70 & 15.27 & 12.47 & 11.73 & 11.27 \cr 
76 & SSSPM J2353$-$4123     $^{  b}$ & 23$:$53$:$01.41 & $-$41$:$23$:$24.6 & 1999.70 &  $+$130 &    $+$0 & 20 &  8 & 22.23 & 19.08 & 16.89 & 14.39 & 13.73 & 13.33 \cr 
77 & APMPM J2354$-$3316     $^{a,b}$ & 23$:$54$:$09.28 & $-$33$:$16$:$26.6 & 1999.57 &  $-$326 &  $-$389 &  8 & 13 & 22.20 & 19.30 & 16.57 & 13.05 & 12.36 & 11.88 \cr 
78 & SSSPM J2356$-$3426     $^{a,b}$ & 23$:$56$:$10.81 & $-$34$:$26$:$04.4 & 1999.57 &   $+$70 &  $-$301 &  9 &  5 & 22.21 & 19.18 & 16.15 & 12.95 & 12.38 & 11.97 \cr 
79 & SSSPM J2400$-$2008     $^{   }$ & 23$:$59$:$57.62 & $-$20$:$07$:$39.4 & 1998.61 &  $+$402 &  $-$511 & 29 & 10 &        & 20.29 & 17.45 & 14.38 & 13.62 & 13.25 \cr 
\hline
 \end{tabular}
 \label{tablecolours}
\end{center}
\end{table*}

%
\begin{table*}
\thispagestyle{headings}
\begin{center}
 \caption[]{\normalsize
Observing logs for the subset of proper motion objects observed photometrically 
with FORS1 ($R$ and $I$) and ISAAC ($J_s$, $H$, $K_s$) mounted on the VLT\@.
Names, observing dates, and exposure times for the optical and 
near-infrared observations are given for each individual object.
}
\scriptsize
 \begin{tabular}{|rl|ccccc|ccccc|}
 \hline
RN & Name    & \multicolumn{5}{c|}{Optical imaging} & \multicolumn{5}{c|}{Near-infrared imaging} \cr
 \hline
    &    & Date & ExpT($R$) & $R$ & ExpT($I$) & $I$ & Date & ExpT & $J_s$ & $H$ & $K_s$ \cr
 \hline
02 & BRI B0021$-$0214      & 1999 Jun 01 & 3$\times$10s & 17.42 & 3$\times$5s & 15.34 & 1999 Jul 22 & 3$\times$2s & 11.87 & 11.16 & 10.60 \cr
05 & LP 645-52           & 1999 Jun 01 & 3$\times$10s & 17.56 & 3$\times$5s & 15.73 & 1999 Jun 02 & 3$\times$2s & 13.80 & 13.35 & 13.05 \cr
06 & APMPM J0057$-$7604    & 2000 Jun 21 & 3$\times$10s & 16.63 & 3$\times$5s & 14.28 & 2000 Jun 19 & 3$\times$2s & 12.71 & 12.09 & 11.75 \cr
13 & LP 769-14           & 1999 Jul 25 & 3$\times$10s & 18.20 & 3$\times$5s & 16.50 & 1999 Jun 02 & 3$\times$2s & 14.62 & 14.05 & 13.74 \cr
15 & APMPM J0207$-$7214    & 1999 Jun 01 & 3$\times$10s & 18.50 & 3$\times$5s & 17.70 & 1999 Jul 23 & 3$\times$2s & 16.42 & 15.95 & 15.75 \cr
16 & APMPM J0207$-$3722    & 1999 Jul 25 & 3$\times$10s & 17.44 & 3$\times$5s & 15.24 & 1999 Jul 23 & 3$\times$2s & 12.48 & 11.86 & 11.44 \cr
21 & APMPM J0232$-$4437    & 2000 Jun 21 & 3$\times$10s & 18.40 & 3$\times$5s & 16.04 & 2000 Jun 19 & 3$\times$2s & 14.65 & 14.13 & 13.79 \cr
22 & APMPM J0244$-$5203    & 2000 Jun 21 & 3$\times$10s & 18.68 & 3$\times$5s & 16.86 & 2000 Jun 22 & 3$\times$2s & 16.11 & 15.59 & 15.26 \cr
25 & APMPMJ0331$-$2349     & 1999 Aug 29 & 3$\times$10s & 18.50 & 3$\times$5s & 17.00 & 1999 Sep 06 & 3$\times$2s & 15.64 & 15.30 & 15.18 \cr
32 & APMPM J0536$-$5358    & 2000 Apr 08 & 3$\times$10s & 18.20 & 3$\times$5s & 16.50 & 2000 Aug 05 & 3$\times$2s & 14.02 & 13.41 & 13.02 \cr
35 & LP 314-67           & 1999 Apr 23 & 3$\times$10s & 18.10 & 3$\times$5s & 17.00 & 2000 May 12 & 3$\times$2s & --- &  ---  &  ---  \cr
37 & LP 614-35           & 2000 Apr 04 & 3$\times$10s & 18.47 & 3$\times$5s & 17.42 & 1999 May 29 & 3$\times$2s & 16.19 & 15.58 & 15.39 \cr
38 & APMPM J1212$-$2126    & 2000 Apr 04 & 3$\times$10s & 18.41 & 3$\times$5s & 17.36 & 2000 May 15 & 3$\times$2s & 16.26 & 15.81 & 15.69 \cr
39 & APMPM J1222$-$2452    & 2000 Apr 04 & 3$\times$10s & 18.17 & 3$\times$5s & 16.31 & 2000 May 15 & 3$\times$2s & 14.38 & 13.94 & 13.64 \cr
40 & LHS 2555a           & 2000 Apr 04 & 3$\times$10s & 16.92 & 3$\times$5s & 15.64 & 2000 May 15 & 3$\times$2s & 14.37 & 13.91 & 13.70 \cr
41 & APMPM J1251$-$2121    & 2000 Apr 04 & 3$\times$10s & 16.07 & 3$\times$5s & 13.95 & 2000 May 15 & 3$\times$2s & 11.25 & 10.67 & 10.29 \cr
42 & Kelu 1              & 1999 Apr 23 & 3$\times$10s & 19.10 & 3$\times$5s & 17.20 & 2000 Oct 25 & 3$\times$2s & 13.29 & 12.42 & 11.73 \cr
43 & CE 303              & 2000 Apr 02 & 3$\times$10s & 17.00 & 3$\times$5s & 14.77 & 2000 May 15 & 3$\times$2s & 11.88 & 11.31 & 10.88 \cr
44 & CE 352              & 1999 Apr 23 & 3$\times$10s & 18.10 & 3$\times$5s & 17.10 & 1999 Apr 30 & 3$\times$2s & 15.77 & 15.29 & 15.10 \cr
45 & LP 859-1            & 2000 Apr 02 & 3$\times$10s & 16.96 & 3$\times$5s & 14.77 & 2000 Apr 24 & 3$\times$2s & 12.16 & 11.58 & 11.16 \cr
46 & LHS 3141B           & 2000 Apr 02 & 3$\times$10s & 18.46 & 3$\times$5s & 16.15 & 2000 Apr 24 & 3$\times$2s & 13.71 & 13.26 & 13.12 \cr
49 & LP 815-21           & 1999 May 31 & 3$\times$10s & 18.11 & 3$\times$5s & 17.37 & 2000 Oct 25 & 3$\times$2s & 16.12 & 15.64 & 15.39 \cr
51 & APMPM J2036$-$4936    & 2000 Apr 02 & 3$\times$10s & 18.51 & 3$\times$5s & 16.53 & 2000 May 15 & 3$\times$2s & 14.72 & 14.29 & 13.98 \cr
56 & [HB88] M12          & 1999 May 31 & 3$\times$10s & 18.76 & 3$\times$5s & 16.58 & 2000 Oct 25 & 3$\times$2s & 14.51 & 13.92 & 13.58 \cr
57 & LP 819-9            & 1999 May 31 & 3$\times$10s & 18.10 & 3$\times$5s & 16.65 & 1999 May 29 & 3$\times$2s & 15.05 & 14.60 & 14.32 \cr
60 & LDS4980 B           & 1999 May 31 & 3$\times$10s & 18.33 & 3$\times$5s & 16.75 & 1999 Jun 02 & 3$\times$2s & 14.66 & 14.16 & 13.88 \cr
61 & LDS4980 A           & 1999 May 31 & 3$\times$40s & 18.09 & 3$\times$10s & 16.35 & 2000 Oct 25 & 3$\times$2s &  ---  &  ---  &  ---  \cr
69 & APMPM J2330$-$4737    & 1999 Jun 01 & 3$\times$40s & 15.91 & 3$\times$10s & 13.78 & 2000 Sep 22 & 3$\times$2s & 11.27 & 10.73 & 10.37 \cr
70 & APMPM J2331$-$2750    & 1999 Jun 01 & 3$\times$40s & 16.68 & 3$\times$10s & 14.45 & 2000 Oct 25 & 3$\times$2s & 11.67 & 11.09 & 10.73 \cr
72 & APMPM J2344$-$2906    & 1999 Jun 01 & 3$\times$40s & 17.96 & 3$\times$10s & 15.71 & 1999 Jul 22 & 3$\times$2s & 13.19 & 12.78 & 12.42 \cr
74 & APMPM J2347$-$3154    & 2000 May 12 & 3$\times$40s & 18.45 & 3$\times$10s & 16.20 & 2000 Jun 19 & 3$\times$2s & 13.75 & 12.82 & 12.34 \cr
77 & APMPM J2354$-$3316    & 2000 May 12 & 3$\times$40s & 18.36 & 3$\times$10s & 16.15 & 2000 May 15 & 3$\times$2s & 13.21 & 12.61 & 12.10 \cr
\hline
 \end{tabular}
 \label{table_img_OBs}
\end{center}
\end{table*}

%
%
\begin{table*}
\thispagestyle{headings}
\begin{center}
 \caption[]{\normalsize
Observing logs for the first set of proper motion objects observed 
spectroscopically with VLT/FORS1 and VLT/ISAAC in the optical and
near-infrared, respectively. Names, observing dates, and exposure times 
are provided for both wavelength ranges. All optical spectra were obtained 
with Grism\_300I except for APMPM\,J2036-4936 and LP\,859-1 which were
observed with Grism\_150I\@. The near-infrared spectra of the two optically 
classified extreme subdwarfs LP\,815-21 and CE\,352 were noisy and unusable, 
and therefore are not shown in Fig.\  \ref{specIR}.
}
 \footnotesize
 \begin{tabular}{|r@{\hspace{2mm}}l|cc|c@{\hspace{2mm}}c@{\hspace{2mm}}c@{\hspace{2mm}}c|}
 \hline
RN & Name & \multicolumn{2}{c|}{Optical spectroscopy} & \multicolumn{4}{c|}{Near-infrared spectroscopy} \cr
\hline
   &      &  Date & ExpT & Date & ExpT(1.10--1.39$\mu$m) & ExpT(1.42--1.83$\mu$m) & ExpT(1.84--2.56$\mu$m) \cr
 \hline
02 & BRI B0021$-$0214     &  1999 Aug 31 & 3$\times$100s & --- & --- & --- & --- \cr
05 & LP 645-52            &  1999 Aug 31 & 3$\times$100s & --- & --- & --- & --- \cr
13 & LP 769-14            &  2000 Jun 21 & 5$\times$100s & --- & --- & --- & --- \cr
16 & APMPM J0207$-$3722   &  1999 Aug 29 & 3$\times$100s & --- & --- & --- & --- \cr
35 & LP 314-67            &  2000 Apr 05 & 3$\times$100s & --- & --- & --- & --- \cr
37 & LP 614-35            &  2000 Apr 04 & 3$\times$250s & --- & --- & --- & ---  \cr
39 & APMPM J1222$-$2452   &  2000 Jun 21 & 5$\times$100s & --- & --- & --- & ---  \cr
41 & APMPM J1251$-$2121   &  2000 Jun 21 & 3$\times$100s & --- & --- & --- & --- \cr
42 & Kelu 1               &  1999 Jul 07 & 3$\times$250s & 1999 Jul 27 & 4$\times$50s & 4$\times$50s & 6$\times$50s \cr
43 & CE 303               &  2000 Jun 21 & 3$\times$100s & --- & --- & --- & --- \cr
44 & CE 352               &  1999 Jul 07 & 3$\times$100s & 1999 Jul 28 & 10$\times$100s & 10$\times$100s & 18$\times$100s \cr
45 & LP 859-1             &  2000 Apr 24 & 3$\times$100s & --- & --- & --- & ---  \cr
46 & LHS 3141B            &  2000 May 06 & 5$\times$250s & --- & --- & --- & ---  \cr
49 & LP 815-21            &  1999 Aug 27 & 3$\times$250s & 1999 Jul 06 & 5$\times$100s & 5$\times$100s & 18$\times$100s \cr
51 & APMPM J2036$-$4936   &  1999 Aug 27 & 3$\times$250s & --- & --- & --- & ---  \cr
56 & [HB88] M12           &  1999 Aug 26 & 3$\times$250s & 1999 Jul 06 & 4$\times$100s & 4$\times$100s & 6$\times$100s \cr
57 & LP 819-9             &  1999 Aug 29 & 3$\times$250s & --- & --- & --- & ---  \cr
60 & LDS4980 B            &  1999 Aug 29 & 3$\times$250s & --- & --- & --- & ---  \cr
61 & LDS4980 A            &  1999 Aug 29 & 3$\times$250s & --- & --- & --- & ---  \cr
69 & APMPM J2330$-$4737   &  1999 Aug 29 & 5$\times$100s & --- & --- & --- & ---  \cr
70 & APMPM J2331$-$2750   &  1999 Aug 28 & 6$\times$100s & 1999 Jul 23 & 4$\times$25s & 4$\times$25s & 6$\times$25s \cr
72 & APMPM J2344$-$2906   &  1999 Aug 28 & 5$\times$100s & 1999 Jul 23 & 4$\times$25s & 4$\times$25s & 6$\times$25s \cr
\hline
 \end{tabular}
 \label{table_OBs_spec}
\end{center}
\end{table*}

%
%
\begin{center}
\begin{figure*}
\includegraphics[width=18.5cm, angle=0]{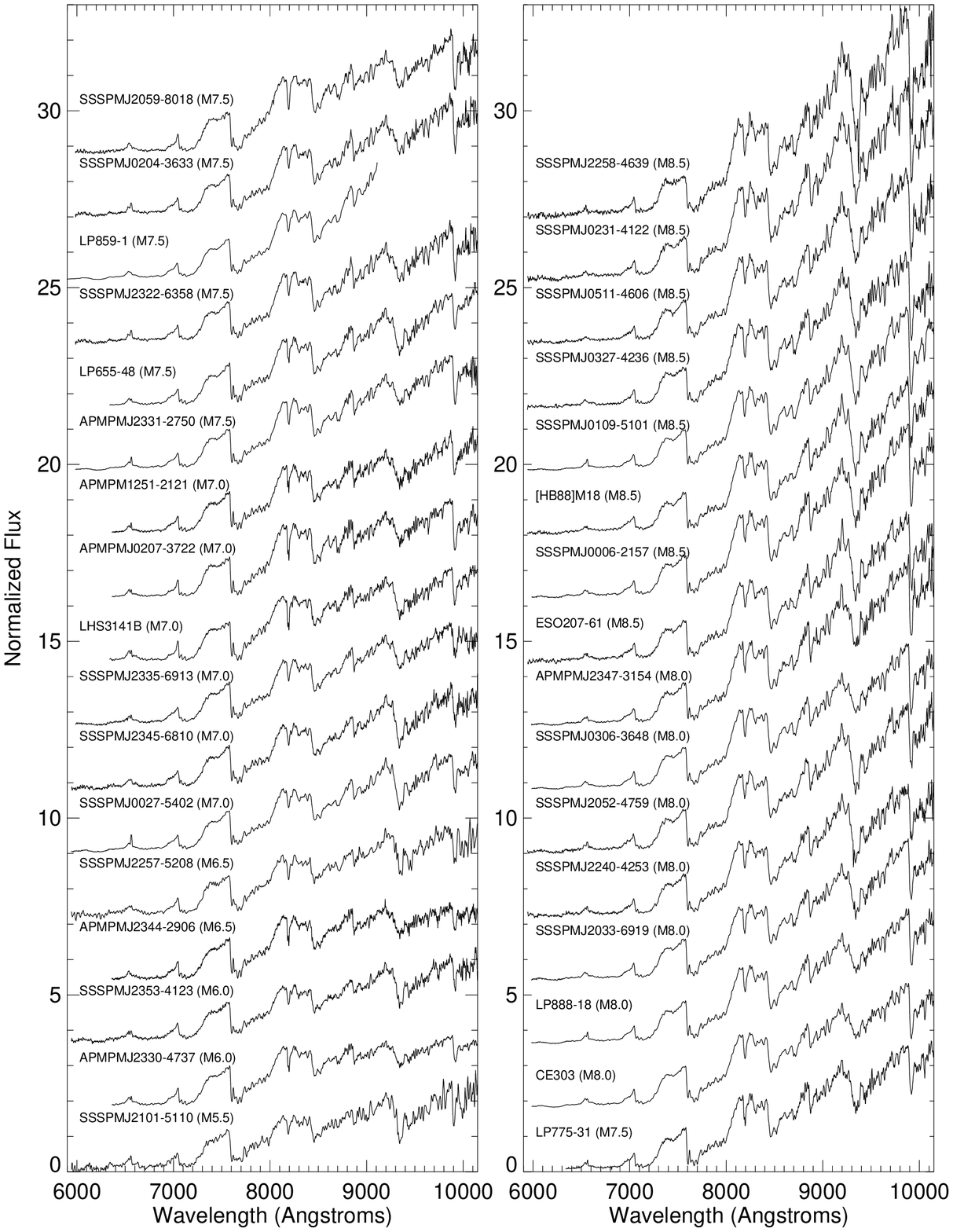}
\caption[Low-resolution spectra of proper motions objects
spanning M5.5--M8.5]{\normalsize
Optical spectra of sources with spectral types between M5.5 and M8,
assigned according to the Kirkpatrick et al.\ (\cite{kirkpatrick99})
and Mart\'{\i}n et al.\ (\cite{martin99}) classification schemes.
The uncertainty on the spectral type is typically half a subclass.
An arbitrary constant has been added to separate the spectra.
}
\label{laterM5}
\end{figure*}
\end{center}

\subsubsection{VLT/FORS1 spectroscopy} 
\label{FORS1_Spectro}
\begin{center}
\begin{figure}[htbp]
\includegraphics[width=\linewidth, angle=0]{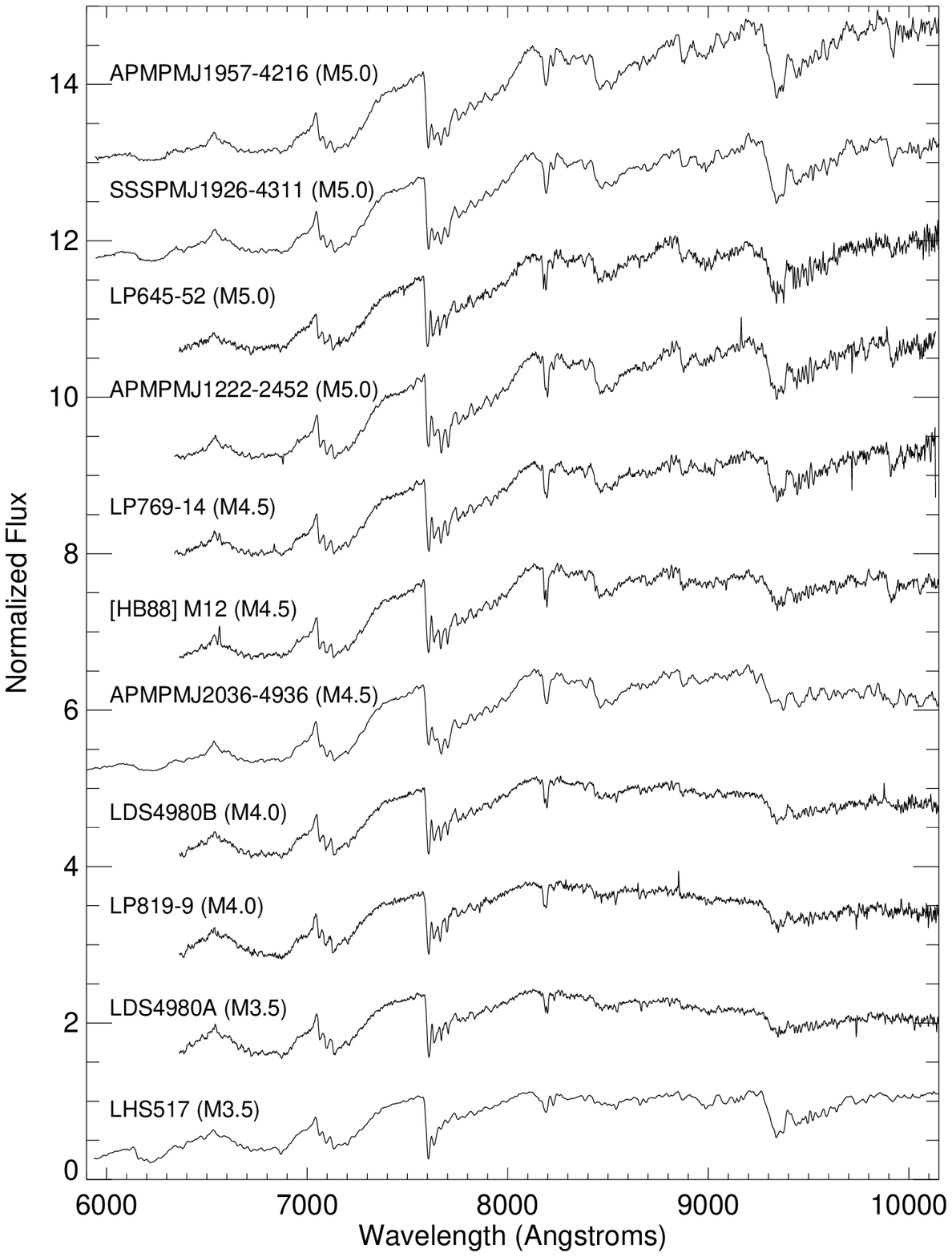}
\caption[Spectra of proper motions objects earlier than M5.5]{\normalsize
Optical spectra of objects earlier than M5.5
according to the Kirkpatrick et al.\ (\cite{kirkpatrick99})
and Mart\'{\i}n et al.\ (\cite{martin99}) classification schemes.
The uncertainty on the spectral type is typically half a subclass.
An arbitrary constant has been added to separate the spectra.
}
\label{belowM5}
\end{figure}
\end{center}

Of the 32 sources observed photometrically with FORS1, a subset of 22
were followed-up with low-resolution FORS1 spectroscopy
(R$\sim$600) with grism GRIS\_300I+11 spanning 
6000--11000\AA{}, apart from APMPM\,J2036$-$4936 and LP\,859-1 which were
observed with grism GRIS\_150I+17 covering 3500--11000\AA{}. Total exposure 
times were set according to the brightness of the object as listed in
Table \ref{table_OBs_spec}. Data reduction involved subtracting an averaged 
dark frame and dividing by a dome flat field. Wavelength calibration was
carried out using He and Ar arc lamp lines covering the whole wavelength range. 
The data were flux calibrated using an averaged sensitivity
function created from a spectrophotometric standard star observed on the
same night. The spectra shown in Figs.\ \ref{belowM5} and \ref{laterM5}
have been normalised at 7500\AA{} for consistency with existing spectral
classification schemes (Kirkpatrick et al.\ \cite{kirkpatrick99};
Mart\'{\i}n et al.\ \cite{martin99}). No telluric correction was
applied to the optical spectra.

%
%

\subsubsection{VLT/ISAAC imaging}
\label{ISAAC_Imaging}
All but two (LP\,314-67 and LDS\,4980A) of the sources imaged with FORS1 
were then imaged with ISAAC to obtain near-infrared photometry.
ISAAC employs a HAWAII 1024$\times$1024 pixel HgCdTe array covering the
1--2.5\,$\mu$m wavelength range with a pixel size of 0.147 arcsec, yielding
a field of view of 2.5$\times$2.5 arcmin. A series of five dithered positions
was observed in each of three broad-band filters ($J_s$, $H$, $K_s$): for
each position in each filter, a flat-field correction was applied and
an average of the other four frames was used to subtract the sky flux.
Aperture photometry was carried out for each individual filter. The
magnitudes listed in Table \ref{table_img_OBs} correspond to the average
of all five measurements in each filter. 
It should be noted that the ISAAC $J_s$ filter is non-standard, thus 
potentially leading to problems when comparing with existing calibrations. 
However, although it was not available when we began our observational program,
the 2MASS Second Incremental Data Release in March 2000 provided standard $J$
magnitudes for our entire sample, and it is these magnitudes which are
in Table \ref{tablecolours} and are used for all calculations made in this paper.

\subsubsection{VLT/ISAAC spectroscopy}
\label{ISAAC_Spectro}
Low-resolution (R\,$\sim$\,600) spectroscopy spanning 1--2.5\,$\mu$m was 
obtained with ISAAC for 6 objects (see Table \ref{table_OBs_spec}) including 
Kelu\,1 (Ruiz et al.\ \cite{ruiz97}), which was observed as a template. A slit 
width of 1 arcsec was employed throughout and three gratings were used to 
cover the spectral ranges 1.10--1.39\,$\mu$m, 1.42--1.83\,$\mu$m, and
1.84--2.56\,$\mu$m. The sources were measured at three or more positions 
along the slit to permit removal of the sky spectrum. Featureless standards
(typically F and G stars) were measured within one degree on the sky of each
source and close in time in order to remove telluric absorption. Each 
individual frame was flat-fielded, sky-subtracted, and a one-dimensional 
spectrum extracted. These spectra were then divided by 
the standard, multiplied by an appropriate spectral template smoothed to our 
resolution, and then co-added to increase the signal-to-noise. The spectra as 
shown in Fig.~\ref{specIR} were normalised at 1.28\,$\mu$m.

%
%
\begin{center}
\begin{figure}[!b]
\includegraphics[width=\linewidth, angle=0]{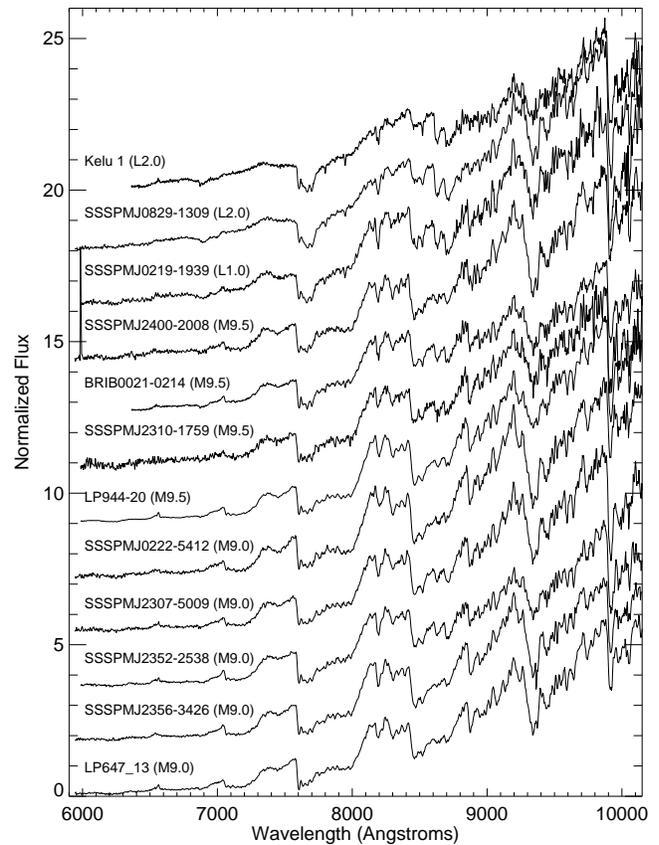}
\caption[Optical spectroscopy of our latest M dwarfs and early
L dwarfs]{\normalsize
Optical spectra of the latest-type (M9 to L2) objects, following
the Kirkpatrick et al.\ (\cite{kirkpatrick99})
and Mart\'{\i}n et al.\ (\cite{martin99}) classification schemes.
The uncertainty on the spectral type is typically half a subclass.
An arbitrary constant has been added to separate the spectra.
}
\label{latest}
\end{figure}
\end{center}

\subsection{ESO\,3.6-m/EFOSC2 spectroscopy}
\label{EFOSC2_Spectro}
A second subsample of 47 very red, high proper motion sources was observed
in the optical with EFOSC2 on the ESO 3.6-m telescope on La Silla on 22--23
November 2001 and 5--8 December 2002. The weather during the 2001 run was
photometric, with seeing $\sim$0.6--0.8 arcsec FWHM; the conditions during
the 2002 run were poorer, with some observations affected by thin clouds
and seeing of $\sim$1.0--1.5 arcsec FWHM\@.

\begin{figure}[!h]
\begin{center}
\includegraphics[width=9.0cm, angle=0]{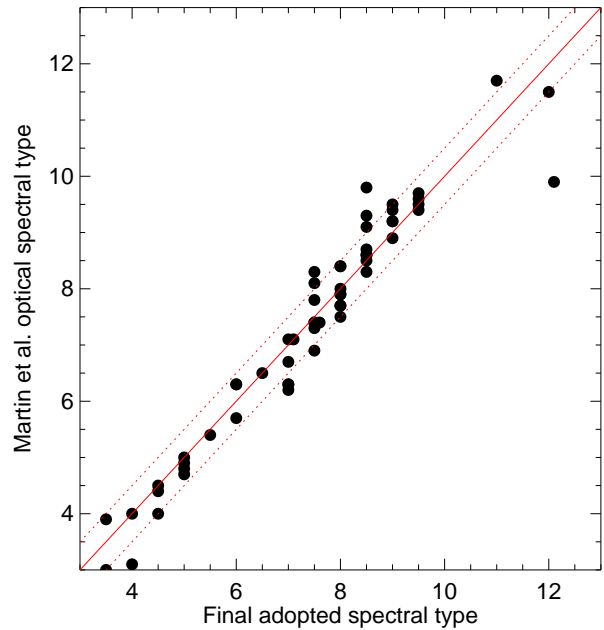}
\caption{
The final optical spectral types (4\,$\equiv$\,M4, \ldots{}, 12\,$\equiv$\,L2)
adopted for each source in this paper vs.\ spectral types derived from
the classification scheme of Mart\'{\i}n et al.\ (\cite{martin99}).
The solid line represents equal spectral types, while the dashed lines
show differences of half a spectral type.
Some sources have been shifted slightly on the x-axis for clarity.
}
\label{M99vsfin}
\end{center}
\end{figure}

\begin{figure}[!h]
\begin{center}
\includegraphics[width=9.0cm, angle=0]{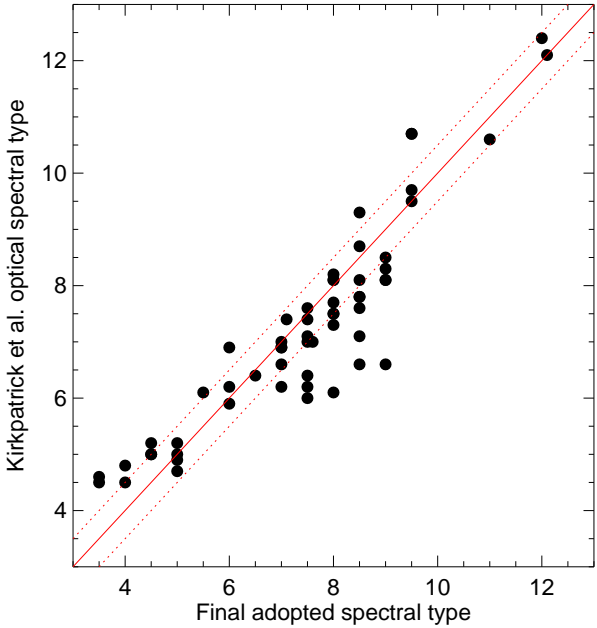}
\caption{
The final optical spectral types (4\,$\equiv$\,M4, \ldots{}, 12\,$\equiv$\,L2)
adopted for each source in this paper vs.\ spectral types derived from
the classification scheme of Kirkpatrick et al.\ (\cite{kirkpatrick99}).
Some sources have been shifted slightly on the x-axis for clarity.
}
\label{K99vsfin}
\end{center}
\end{figure}

\begin{figure}[!h]
\begin{center}
\includegraphics[width=9.0cm, angle=0]{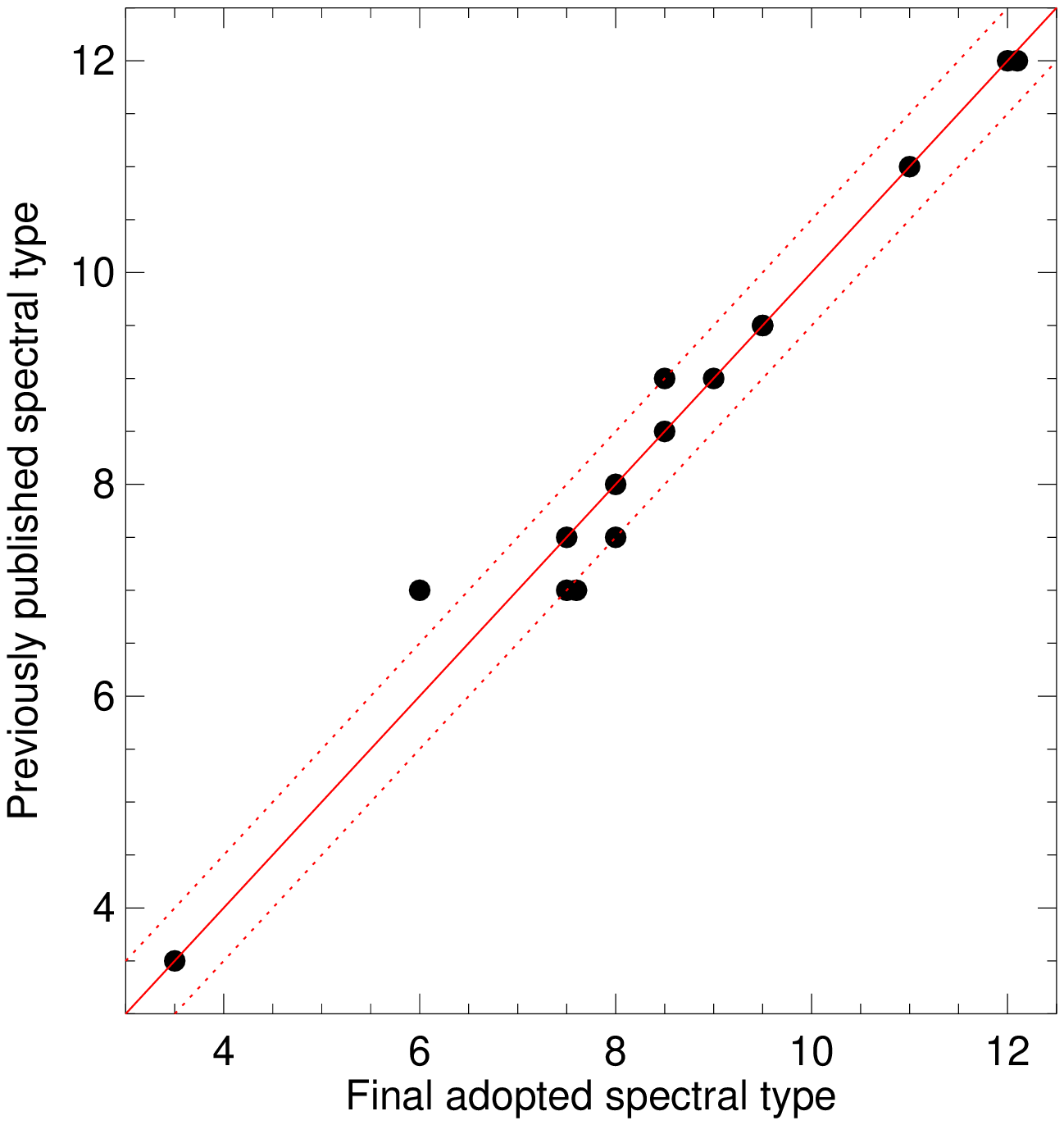}
\caption{
The adopted optical spectral type vs.\ previous classifications found in the
literature (4\,$\equiv$\,M4, \ldots{}, 12\,$\equiv$\,L2).
The latter were obtained from:
Kirkpatrick et al.\ (\cite{kirkpatrick00}): BRI B0021-0214 (M9.5);
Ruiz et al.\ \cite{ruiz97}: Kelu 1 (L2.0);
Tinney \cite{tinney98}: LP\,944-20 (M9.5);
Reid et al.\ (\cite{reid02c}): LP\,859-1 (M7.5);
Gizis et al.\ (\cite{gizis00}): 2MASSW\,J0952-19 (M7.0);
Cruz \& Reid (\cite{cruz02}): LP\,647-13 (M9.0);
Ianna \& Fredrick (\cite{ianna95}): ESO\,207-61 (M9.0);
Scholz \& Meusinger (\cite{scholz02b}): SSSPM\,J0829-1309 (L2.0);
Hawkins \& Bessell (\cite{hawkins88}): [HB88]\,M18 (M8.5);
Hawley, Gizis \& Reid (\cite{hawley96}): LHS\,517 (M3.5);
Cruz et al.\ (\cite{cruz03}): SSSPM\,J0219-1939 (L1.0), LP\,888-18 (M7.5),
LP\,775-31 (M7.0), LP\,655-48 (M7.0), and CE\,303 (M8.0).
2MASSW J0952-19 is the only object with more than half a spectral type
difference in its classification.
Some sources have been
shifted slightly on the x-axis for clarity.
}
\label{SpTcomp}
\end{center}
\end{figure}

EFOSC2 uses a 2048$\times$2048 pixel Loral/Lesser CCD with a pixel size of 
0.157 arcsec, yielding a useful field of view of 5.2$\times$5.2 arcmin.
Spectroscopy was obtained for most of the objects just once, but additional
data were taken for potentially interesting candidates or targets with poorer
signal-to-noise. In these latter cases, the sources were observed at two or 
three positions shifted along the slit by 100 pixels. The details for the
observations for each source are given in Table \ref{table_img_OBs_01}. The 
data were reduced in a similar fashion to that described for the VLT/FORS1 
spectroscopy, with the addition that an internal quartz flat field taken 
immediately after the first exposure in each series was used to remove 
fringing above 8000\AA{}\@. The spectra were normalised at 7500\AA{} and 
are displayed in Figs.\ \ref{latest}--\ref{subdw}. No telluric absorption
correction was applied to the spectra.

%
%
\begin{table*}[htbp]
\thispagestyle{headings}
\begin{center}
 \caption[]{\normalsize
Observing logs for all proper motion targets observed in the optical with
Grism\#12 in EFOSC2 on the ESO 3.6-m and in the near-infrared with the Blue
and Red gratings in SOFI on the NTT\@. Names, observing dates, and exposure
times are provided for both wavelength ranges.
}
 \small
 \begin{tabular}{|rl|cc|cc|}
 \hline
RN & Name   & \multicolumn{2}{c|}{Optical spectroscopy} & \multicolumn{2}{c|}{Near-infrared spectroscopy} \cr
\hline
   &   &  Date & ExpT & Date & ExpT \cr
 \hline
01 & SSSPM J0006$-$2157       & 2001 Nov 21 & 3$\times$1000s & --- & --- \cr
03 & SSSPM J0027$-$5402       & 2002 Dec 06 & 1$\times$400s & --- & --- \cr
04 & SSSPM J0030$-$3427       & --- & --- & 2001 Nov 25 & 3$\times$150s \cr
07 & SSSPM J0109$-$5101       & 2002 Dec 05 & 3$\times$400s & 2001 Nov 25 & 3$\times$60s \cr
08 & SSSPM J0109$-$4955       &  --- & --- &  2001 Nov 25 & 3$\times$150s \cr
09 & LP 647-13                & 2002 Dec 05 & 1$\times$360s & --- & --- \cr
10 & SSSPM J0124$-$4240       & --- & --- & 2001 Nov 25 & 3$\times$150s \cr
11 & SSSPM J0125$-$6546       & --- & ---& 2001 Nov 25 & 3$\times$200s \cr
12 & SSSPM J0134$-$6315       & --- & --- & 2001 Nov 25 & 3$\times$150s \cr
14 & SSSPM J0204$-$3633       & 2001 Nov 22 & 1$\times$500s & 2001 Nov 24 & 3$\times$200s \cr
16 & APMPM J0207$-$3722       & --- & --- & 2001 Nov 24 & 3$\times$150s \cr
17 & SSSPM J0215$-$4804       & --- & --- & 2001 Nov 25 & 3$\times$150s \cr
18 & SSSPM J0219$-$1939       & 2001 Nov 21 & 3$\times$900s & 2001 Nov 24 & 3$\times$250s \cr
19 & SSSPM J0222$-$5412       & 2002 Dec 08 & 2$\times$900s & --- & --- \cr
20 & SSSPM J0231$-$4122       & 2002 Dec 08 & 1$\times$800s & --- & --- \cr
23 & SSSPM J0306$-$3648       & 2001 Nov 22 & 1$\times$300s & 2001 Nov 24 & 3$\times$120s \cr
24 & SSSPM J0327$-$4236       & 2002 Dec 06 & 1$\times$900s & --- & --- \cr
26 & LP 888-18                & 2001 Nov 22 & 1$\times$270s & 2001 Nov 24 & 3$\times$100s \cr
27 & LP 944-20                 & 2001 Nov 22 & 1$\times$240s & 2001 Nov 24 & 3$\times$60s \cr   
28 & LP 775-31                 & 2001 Nov 22 & 1$\times$200s & 2001 Nov 24 & 3$\times$60s \cr   
29 & LP 655-48                 & 2001 Nov 22 & 1$\times$200s & 2001 Nov 24 & 3$\times$60s \cr   
30 & SSSPM J0500$-$5406       & 2001 Nov 22 & 1$\times$300s & --- & --- \cr
31 & SSSPM J0511$-$4606       & 2002 Dec 08 & 1$\times$780s & --- & --- \cr
33 & ESO 207-61               & 2002 Dec 06 & 2$\times$600s & --- & --- \cr
34 & SSSPM J0829$-$1309       & 2002 Dec 08 & 1$\times$480s & --- & --- \cr
36 & 2MASS J0952$-$19         & 2001 Nov 22 & 1$\times$240s & --- & --- \cr
47 & SSSPM J1926$-$4311       & 2002 Dec 05 & 1$\times$240s & --- & --- \cr
48 & APMPM J1957$-$4216       & 2002 Dec 06 & 1$\times$240s & --- & --- \cr
50 & SSSPM J2033$-$6919       & 2001 Nov 21 & 3$\times$1200s & --- & --- \cr
52 & SSSPM J2052$-$4759       & 2002 Dec 05 & 1$\times$540s & --- & --- \cr
53 & SSSPM J2059$-$8018       & 2001 Nov 22 & 2$\times$900s & --- & --- \cr
55 & [HB88] M18               & 2002 Dec 06 & 2$\times$600s & --- & --- \cr
58 & LHS 517                  & 2002 Dec 08 & 1$\times$15s & --- & --- \cr
59 & SSSPM J2229$-$6931       & --- & --- & 2001 Nov 25 & 3$\times$200s \cr
62 & SSSPM J2240$-$4253       & 2002 Dec 06 & 1$\times$660s & --- & --- \cr
63 & SSSPM J2257$-$5208       & 2002 Dec 08 & 1$\times$720s & --- & --- \cr
64 & SSSPM J2258$-$4639       & 2002 Dec 08 & 1$\times$720s & --- & --- \cr
65 & SSSPM J2307$-$5009       & 2002 Dec 06 & 1$\times$660s & 2001 Nov 25 & 3$\times$200s \cr
66 & SSSPM J2310$-$1759       & 2001 Nov 22 & 3$\times$900s & 2001 Nov 24 & 3$\times$250s \cr
67 & SSSPM J2319$-$4919       & --- & --- & 2001 Nov 25 & 3$\times$150\,s \cr
68 & SSSPM J2322$-$6358       & 2001 Nov 22 & 1$\times$600s & --- & --- \cr
71 & SSSPM J2335$-$6913       & 2001 Nov 22 & 1$\times$600s & --- & --- \cr
73 & SSSPM J2345$-$6810       & 2002 Dec 08 & 1$\times$600s & 2001 Nov 25 & 3$\times$200s  \cr
74 & APMPM J2347$-$3154       & 2001 Nov 21 & 3$\times$1200s & --- & --- \cr
75 & SSSPM J2352$-$2538       & 2001 Nov 22 & 1$\times$400s & 2001 Nov 24 & 3$\times$150s \cr
76 & SSSPM J2353$-$4123       & 2002 Dec 06 & 1$\times$540s & --- & --- \cr
77 & APMPM J2354$-$3316       & 2002 Dec 08 & 1$\times$540s & 2001 Nov 25 & 3$\times$150s \cr
78 & SSSPM J2356$-$3426       & 2002 Dec 08 & 2$\times$500s & 2001 Nov 25 & 3$\times$150s \cr
79 & SSSPM J2400$-$2008       & 2002 Dec 08 & 3$\times$900s & 2001 Nov 24 & 3$\times$250s \cr
\hline
 \end{tabular}
 \label{table_img_OBs_01}
\end{center}
\end{table*}

\subsection{NTT/SOFI spectroscopy}
\label{NTT_Spectro}
Additional near-infrared (0.95--2.5\,$\mu$m) spectroscopy was obtained using
SOFI on the ESO NTT on La Silla on 24--25 November 2001 for sources identified
for follow-up from the EFOSC2 optical spectroscopy and for sources with
interesting near-infrared colours in the 2MASS database. Both nights were
photometric with seeing varying between 0.8--1.0 arcsec FWHM\@.

%
%
\begin{center}
\begin{figure*}[htbp]
\includegraphics[width=18.5cm, angle=0]{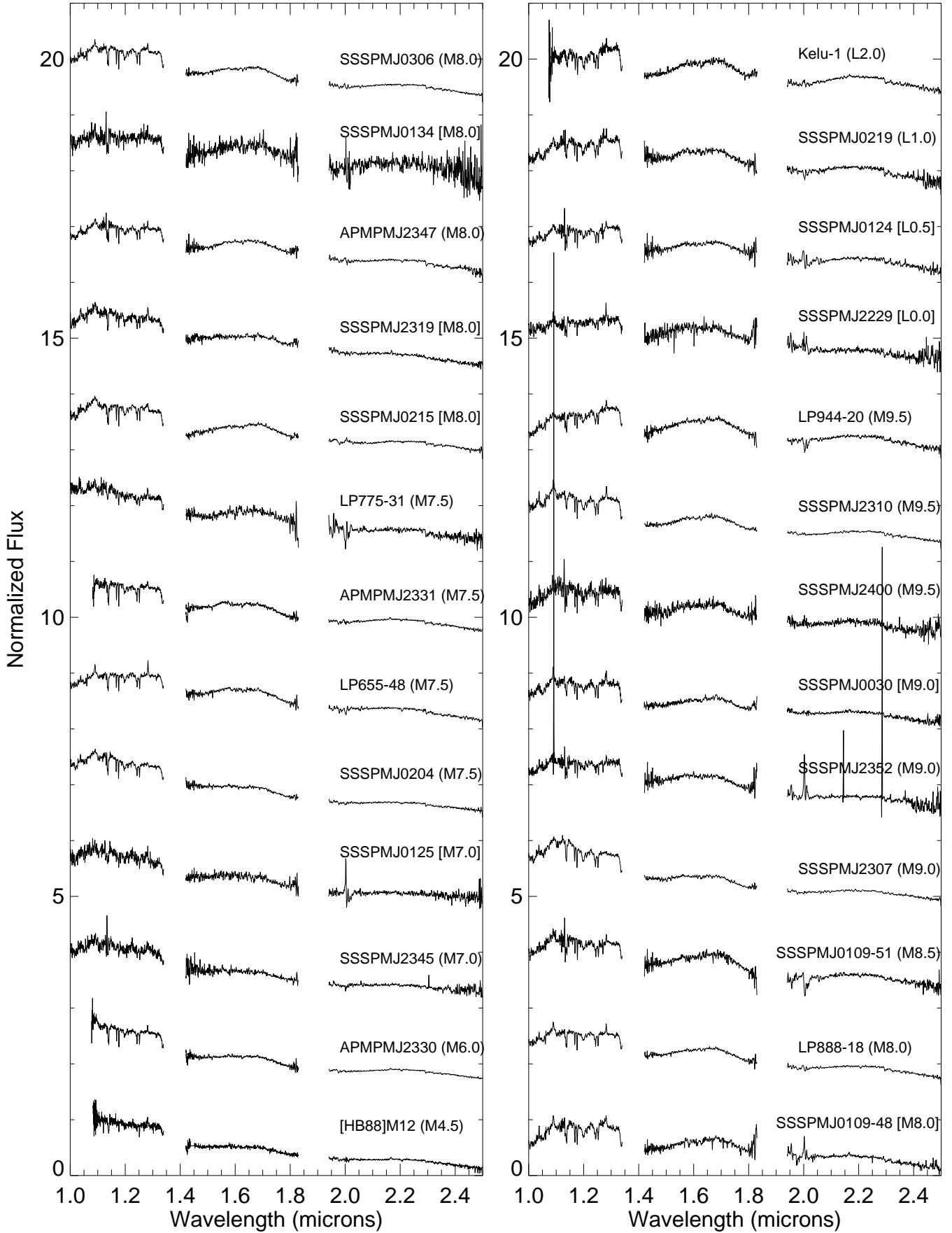}
\caption[Infrared spectra of 26 proper motion objects]{\normalsize
Near-infrared spectra of a subsample of the objects.
Objects with spectral types in square brackets have been
classified based on their near-infrared spectra alone, while
those listed in parentheses are derived from the optical
spectrum for each source.
An arbitrary constant has been added to separate the spectra.
}
\label{specIR}
\end{figure*}
\end{center}

SOFI employs a HAWAII 1024$\times$1024 pixel HgCdTe array (Moorwood \&
Spyromilio \cite{moorwood}) with a pixel size of 0.294 arcsec delivered by the
Large Field Objective used for spectroscopy.  A 1 arcsec slit was used yielding
R$\sim$600 for both blue (0.95--1.64\,$\mu$m) and red (1.53--2.52 $\mu$m) 
gratings. Featureless standards (typically F5--F7) were measured within one
degree on the sky of each source to remove telluric absorption and flux 
calibrate. Each source was observed at three positions along the slit to
facilitate sky subtraction. The data were reduced in the same way as 
described in \S\ref{ISAAC_Spectro}. The spectra were again normalised at 
1.28\,$\mu$m and are shown in Fig.\ \ref{specIR}.

%
%
\begin{table*}[!htbp]
\begin{center}
\caption{
\normalsize Results from the optical spectra for all M and L
dwarfs in our sample. The measured H$\alpha$ index and equivalent width
as discussed in the text are listed first. Then optical indices (TiO5, VO-a,
and PC3) and spectral types in the schemes of Kirkpatrick et al.\
(\cite{kirkpatrick99}) and Mart\'{\i}n et al.\ (\cite{martin99}) are
shown. SpT (TiO5), SpT (VO-a), and SpT (PC3) show
the numerical values and corresponding spectral type for the three
indices, while SpT (Comp) shows the spectral type assigned via
direct comparison with template objects. The final, adopted spectral type
corresponds to the mean value of the three spectral indices rounded to the
nearest half spectral type or to that obtained from direct comparison as
discussed in the text. The typical uncertainty of the adopted spectral type
is half a subclass. The distance shown is derived from the relation between
the absolute M$_J$ magnitude and the spectral type presented in
Dahn et al.\ (\cite{dahn02}); $v_t$ is the resulting tangential velocity.
}
 \scriptsize
\begin{tabular}{@{\hspace{3mm}}r@{\hspace{3mm}}l@{\hspace{3mm}}c@{\hspace{3mm}}r@{\hspace{3mm}}c@{\hspace{3mm}}c@{\hspace{3mm}}c@{\hspace{3mm}}c@{\hspace{3mm}}l@{\hspace{3mm}}c@{\hspace{3mm}}c}
\hline
\hline
\smallskip
RN & Name     &   H$\alpha$  & H${\alpha}$ & SpT  & SpT  & SpT & SpT  & SpT   & distance & $v_t$ \cr
   &          &   index      & EW          & TiO5 & VO-a & PC3 & Comp & final & [pc]     & [km/s] \cr
\hline
01 & SSSPM J0006$-$2157       & 1.013 &  1.9  & 0.275 (M7.8)  & 2.290 (M7.8)  &  1.982 (M8.6) & M8.5 & M8.5 & ~25.0\,$\pm$\,3.0 &  87 \cr
02 & BRI B0021$-$0214         & 0.938 &  0.0  & 0.670 (L0.0)  & 2.275 (L0.7)  &  2.344 (M9.7) & M9.5 & M9.5 & ~11.9\,$\pm$\,1.4 &  9 \cr
03 & SSSPM J0027$-$5402       & 1.947 & 14.5  & 0.278 (M7.8)  & 2.213 (M7.0)  &  1.500 (M6.3) & M7.0 & M7.0 & ~20.8\,$\pm$\,2.5 & 41 \cr
05 & LP 645-52\_1             & 0.920 &  0.2  & 0.314 (M4.8)  & 1.996 (M4.7)  &  1.276 (M5.0) & M5.0 & M5.0 & 100.5\,$\pm$\,12.1  & 209 \cr
05 & LP 645-52\_2             & 0.930 &  0.2  & 0.310 (M4.9)  & 2.012 (M4.9)  &  1.228 (M4.6) & M5.0 & M5.0 & 100.5\,$\pm$\,12.1  & 209 \cr
07 & SSSPM J0109$-$5101       & 1.320 & 13.3  & 0.328 (M8.1)  & 2.293 (M7.8)  &  1.996 (M8.6) & M8.5 & M8.5 & ~15.5\,$\pm$\,1.9 & 17 \cr
09 & LP647-13                 & 1.310 & -7.9  & 0.56  (M9.4)  & 2.360 (M8.5)  &  2.080 (M8.9) & M9.0 & M9.0 & ~11.2\,$\pm$\,1.3 & 19 \cr
13 & LP 769-14                & 1.136 &  2.5  & 0.335 (M4.6)  & 2.023 (M5.0)  &  1.204 (M4.5) & M4.5 & M4.5 & 159.2\,$\pm$\,19.1  & 142 \cr
14 & SSSPM J0204$-$3633       & 1.529 & 10.8  & 0.252 (M7.6)  & 2.254 (M7.4)  &  1.701 (M7.4) & M7.5 & M7.5 &  29.3\,$\pm$\,3.5 & 31 \cr
16 & APMPM J0207$-$3722       & 1.507 &  6.7  & 0.210 (M7.4)  & 2.257 (M7.4)  &  1.637 (M7.1) & M7.0 & M7.0 & ~21.6\,$\pm$\,2.6 &  46 \cr
18 & SSSPM J0219$-$1939       & 0.935 &  0.0  & 0.951 (L1.6)  & 2.289 (L0.6)  &  3.016 (L1.7) & L0.5 & L1.0 & ~24.9\,$\pm$\,3.0 &  31 \cr
19 & SSSPM J0222$-$5412       & 1.059 &  0.0  & 0.433 (M8.7)  & 2.643 (M8.1)  &  2.163 (M9.2) & M9.0 & M9.0 & ~30.9\,$\pm$\,3.7 & 16 \cr
20 & SSSPM J0231$-$4122       & 0.779 &  0.0  & 0.414 (M8.6)  & 2.267 (M7.6)  &  2.194 (M9.3) & M8.5 & M8.5 & ~32.7\,$\pm$\,3.9 & 51 \cr
23 & SSSPM J0306$-$3648       & 0.978 &  0.5  & 0.224 (M7.5)  & 2.241 (M7.3)  &  1.924 (M8.4) & M8.0 & M8.0 & ~13.1\,$\pm$\,1.6 &  43 \cr
24 & SSSPM J0327$-$4236       & 0.925 &  0.0  & 0.251 (M7.6)  & 2.226 (M7.1)  &  2.133 (M9.1) & M8.5 & M8.5 & ~38.8\,$\pm$\,4.7 & 51 \cr
26 & LP 888-18                & 1.544 & 11.0  & 0.264 (M7.7)  & 2.259 (M7.5)  &  1.817 (M7.9) & M8.0 & M8.0 & ~11.2\,$\pm$\,1.3 &  21 \cr
27 & LP 944-20                & 1.001 &  0.0  & 0.521 (M9.2)  & 2.455 (M9.5)  &  2.258 (M9.4) & M9.5 & M9.5 & ~~6.6\,$\pm$\,0.8 &  13 \cr
28 & LP775-31                 & 1.183 &  3.7  & 0.257 (M7.7)  & 2.219 (M7.0)  &  1.737 (M7.6) & M7.5 & M7.5 & ~~7.8\,$\pm$\,0.9 &  13 \cr
29 & LP655-48                 & 1.480 & 10.0  & 0.282 (M7.8)  & 2.213 (M7.0)  &  1.666 (M7.2) & M7.5 & M7.5 & ~~8.8\,$\pm$\,1.1 &  15 \cr
31 & SSSPM J0511$-$4606       & 0.796 &  0.0  & 0.311 (M8.0)  & 2.477 (M9.3)  &  2.005 (M8.7) & M8.5 & M8.5 & ~33.3\,$\pm$\,4.0 & 21 \cr
33 & ESO 207-61               & 1.174 &  2.8  & 0.311 (M8.0)  & 2.321 (M8.1)  &  1.969 (M8.5) & M8.5 & M8.5 & ~24.6\,$\pm$\,2.9 & 47 \cr
34 & SSSPM J0829$-$1309       & 0.996 &  0.0  & 1.008 (L1.9)  & 2.033 (L2.4)  &  2.753 (L1.5) & L2.0 & L2.0 & ~11.6\,$\pm$\,1.4 & 33 \cr
36 & 2MASSW J0952$-$19        & 1.575 & 11.3  & 0.240 (M5.6)  & 2.106 (M5.9)  &  1.489 (M6.3) & M6.0 & M6.0 & ~21.0\,$\pm$\,2.5 &  13 \cr
39 & APMPM J1222$-$2452       & 0.867 &  0.0  & 0.287 (M5.1)  & 2.041 (M5.2)  &  1.253 (M4.8) & M5.0 & M5.0 & 129.4\,$\pm$\,15.5  & 279 \cr
41 & APMPM J1251$-$2121       & 1.109 &  2.6  & 0.238 (M7.6)  & 2.272 (M7.6)  &  1.614 (M6.9) & M7.5 & M7.5 & ~11.1\,$\pm$\,1.3 & 29 \cr
42 & Kelu 1\_1                & 0.933 &  0.0  & 0.965 (L1.7)  & 1.983 (L2.8)  &  2.128 (L0.9) & L2.0 & L2.0 & ~15.4\,$\pm$\,1.8 & 24 \cr
42 & Kelu 1\_2                & 1.163 &  0.0  & 1.009 (L1.9)  & 2.087 (L2.1)  &  2.471 (L1.2) & L2.0 & L2.0 & ~15.4\,$\pm$\,1.8 & 24 \cr
43 & CE 303                   & 1.456 &  6.8  & 0.258 (M7.7)  & 2.322 (M8.1)  &  1.771 (M7.7) & M8.0 & M8.0 & ~13.6\,$\pm$\,1.6 & 24 \cr
45 & LP 859-1                 & 0.974 &  0.2  &  ---  (---)   & 2.228 (M7.1)  &  1.697 (M7.4) & M7.5 & M7.5 & ~16.4\,$\pm$\,2.0 & 27 \cr
46 & LHS 3141B                & 1.245 &  4.8  & 0.115 (M6.9)  & 2.059 (M6.9)  &  1.645 (M7.1) & M7.0 & M7.0 & ~37.0\,$\pm$\,4.4 & 101 \cr
47 & SSSPM J1926$-$4311       & 0.912 &  0.0  & 0.303 (M4.9)  & 2.018 (M4.9)  &  1.230 (M4.7) & M5.0 & M5.0 & ~43.0\,$\pm$\,5.2 & 222 \cr
48 & APMPM J1957$-$4216       & 0.966 &  0.0  & 0.264 (M5.4)  & 2.024 (M5.0)  &  1.273 (M4.9) & M5.0 & M5.0 & ~54.4\,$\pm$\,6.5 & 266 \cr
50 & SSSPM J2033$-$6919       & 0.983 &  1.0  & 0.214 (M7.4)  & 2.281 (M7.7)  &  1.803 (M7.9) & M8.0 & M8.0 & ~32.1\,$\pm$\,3.9 &   78 \cr
51 & APMPM J2036$-$4936       & 0.859 &  0.0  & 0.292 (M5.0)  & 2.020 (M5.0)  &  1.140 (M4.0) & M4.0 & M4.5 & 165.8\,$\pm$\,19.9  &  335 \cr
52 & SSSPM J2052$-$4759       & 1.475 & 12.1  & 0.303 (M7.9)  & 2.333 (M8.2)  &  1.847 (M8.0) & M8.0 & M8.0 & ~23.2\,$\pm$\,2.8 &  48 \cr
53 & SSSPM J2059$-$8018       & 0.869 &  0.2  & 0.134 (M6.8)  & 2.123 (M6.0)  &  1.904 (M8.3) & M7.5 & M7.5 & ~46.6\,$\pm$\,5.6 &   81 \cr
54 & SSSPM J2101$-$5110       & 1.212 &  7.1  & 0.299 (M5.0)  & 2.129 (M6.1)  &  1.341 (M5.4) & M6.0 & M5.5 & 107.7\,$\pm$\,12.9  &   72 \cr
55 & [HB88] M18               & 1.044 &  0.0  & 0.364 (M8.3)  & 2.381 (M8.7)  &  1.913 (M8.3) & M8.5 & M8.5 & ~26.9\,$\pm$\,3.2 &  78 \cr
56 & [HB88] M12               & 1.556 &  6.6  & 0.279 (M5.2)  & 2.044 (M5.2)  &  1.190 (M4.4) & M4.5 & M4.5 & 153.0\,$\pm$\,18.4  &    62 \cr
57 & LP 819-9                 & 0.938 &  0.0  & 0.384 (M4.1)  & 2.006 (M4.8)  &  1.134 (M4.0) & M4.0 & M4.0 & 199.8\,$\pm$\,24.0  &  269 \cr
58 & LHS 517                  & 0.956 &  0.0  & 0.506 (M2.7)  & 1.975 (M4.5)  &  1.004 (M3.0) & M3.5 & M3.5 & ~~5.1\,$\pm$\,0.6 &    27 \cr
60 & LDS4980 B                & 0.992 &  0.6  & 0.377 (M4.1)  & 1.984 (M4.6)  &  1.124 (M3.9) & M4.0 & M4.0 & 195.6\,$\pm$\,23.5  &  220 \cr
61 & LDS4980 A                & 0.927 &  0.0  & 0.443 (M3.4)  & 1.978 (M4.5)  &  1.017 (M3.1) & M3.5 & M3.5 & 229.2\,$\pm$\,27.5  &  255 \cr
62 & SSSPM J2240$-$4253       & 1.238 &  7.1  & 0.254 (M7.7)  & 2.257 (M7.5)  &  1.774 (M7.7) & M8.0 & M8.0 & ~33.9\,$\pm$\,4.1 &   87 \cr
63 & SSSPM J2257$-$5208       & 1.006 &  2.6  & 0.196 (M6.1)  & 2.139 (M6.2)  &  1.472 (M6.2) & M7.0 & M7.0 & ~67.9\,$\pm$\,8.1 &   41 \cr
64 & SSSPM J2258$-$4639       & 0.833 &  0.0  & 0.231 (M7.5)  & 2.181 (M6.6)  &  2.430 (M9.8) & M8.5 & M8.5 & ~29.3\,$\pm$\,3.5 &   38 \cr
65 & SSSPM J2307$-$5009       & 1.318 &  2.0  & 0.513 (M9.1)  & 2.334 (M8.3)  &  2.265 (M9.5) & M8.5 & M9.0 & ~24.4\,$\pm$\,2.9 &   53 \cr
66 & SSSPM J2310$-$1759       & 1.550 &  0.0  & 0.520 (M9.2)  & 2.280 (L0.7)  &  2.273 (M9.5) & M9.5 & M9.5 & ~35.7\,$\pm$\,4.3 &    48 \cr
68 & SSSPM J2322$-$6358       & 1.600 & 11.2  & 0.244 (M5.6)  & 2.161 (M6.4)  &  1.677 (M7.3) & M7.5 & M7.5 & ~46.2\,$\pm$\,5.5 &   27 \cr
69 & APMPM J2330$-$4737       & 1.292 &  5.6  & 0.204 (M6.0)  & 2.142 (M6.2)  &  1.495 (M6.3) & M6.0 & M6.0 & ~15.8\,$\pm$\,1.9 &    86 \cr
70 & APMPM J2331$-$2750       & 1.297 &  5.1  & 0.181 (M6.2)  & 2.137 (M6.2)  &  1.850 (M8.1) & M7.5 & M7.5 & ~13.9\,$\pm$\,1.7 &    50 \cr
71 & SSSPM J2335$-$6913       & 1.300 &  6.1  & 0.257 (M7.7)  & 2.173 (M6.6)  &  1.563 (M6.7) & M7.0 & M7.0 & ~42.7\,$\pm$\,5.1 &    37 \cr
72 & APMPM J2344$-$2906       & 0.975 &  3.0  & 0.121 (M6.9)  & 2.161 (M6.4)  &  1.530 (M6.5) & M6.0 & M6.5 & ~34.1\,$\pm$\,4.1 &    64 \cr
73 & SSSPM J2345$-$6810       & 1.034 &  0.0  & 0.240 (M7.6)  & 2.204 (M6.9)  &  1.495 (M6.3) & M7.0 & M7.0 & ~43.9\,$\pm$\,5.3 &    46 \cr
74 & APMPM J2347$-$3154       & 1.094 &  5.6  & 0.263 (M7.7)  & 2.322 (M8.1)  &  1.924 (M8.4) & M8.5 & M8.0 & ~27.2\,$\pm$\,3.3 &    76 \cr
75 & SSSPM J2352$-$2538       & 1.415 &  6.0  & 0.422 (M8.6)  & 2.174 (M6.6)  &  2.221 (M9.4) & M9.0 & M9.0 & ~16.0\,$\pm$\,1.9 &    31 \cr
76 & SSSPM J2353$-$4123       & 1.232 &  2.1  & 0.260 (M7.7)  & 2.204 (M6.9)  &  1.391 (M5.7) & M6.0 & M6.0 & ~65.9\,$\pm$\,7.9 &   41 \cr
77 & APMPM J2354$-$3316       & 2.163 & Flare & 0.774 (L0.6)  & 2.158 (Flare) &  1.475 (M6.2) & M8.5$^{a}$ & M8.5 & ~22.6\,$\pm$\,2.7 &  55 \cr
78 & SSSPM J2356$-$3426       & 1.310 &  0.0  & 0.557 (M9.4)  & 2.359 (M8.5)  &  2.076 (M8.9) & M9.0 & M9.0 & ~20.0\,$\pm$\,2.4 &    29 \cr
79 & SSSPM J2400$-$2008       & 1.048 &  0.0  & 0.612 (M9.7)  & 2.476 (M9.7)  &  2.308 (M9.6) & M9.5 & M9.5 & ~35.7\,$\pm$\,4.3 &   110 \cr
\hline
\end{tabular}
\label{index}
\end{center}
Note: There are two independent data sets for LP\,645-52 and Kelu 1, which are in good agreement. They lead to the same spectral types
and corresponding estimates of the distances and tangential velocities. \\
$^{a}$ The spectral type of this object was derived from the quiet spectrum 
(see Scholz et al.\ \cite{scholz04a} for more details).
\end{table*}

%
%
\section{Spectroscopic classification of M and L dwarfs}
\label{sp_class}
\subsection{Optical classification}

Two independent classification schemes for M and L dwarfs based on their
optical spectra have recently been defined by Kirkpatrick et al.\
(\cite{kirkpatrick99}) and Mart\'{\i}n et al.\ (\cite{martin99}). The former
scheme is based on spectral ratios in different regions of interest and on
spectrum comparisons, while the second scheme is primarily based on
a pseudo-continuum ratio known as PC3. In this paper, we have adopted
classifications based on a hybrid of the most reliable indices for late M
and early L dwarfs, namely TiO5 (Reid, Hawley, \& Gizis \cite{reid95};
Cruz \& Reid \cite{cruz02}),
VO-a (Kirkpatrick et al.\ \cite{kirkpatrick99}; Cruz \& Reid \cite{cruz02}),
and PC3 (Mart\'{\i}n et al.\
\cite{martin99}), as well the direct comparison with objects of known
spectral type from Kirkpatrick et al.\ (\cite{kirkpatrick99}), Gizis et al.\
(\cite{gizis00}), and Reid et al.\ (\cite{reid99}) which we observed with
the same instrument setups as our candidates. We also compared our spectra
with templates available from other groups, including those of Neill
Reid\footnote{http://dept.physics.upenn.edu/$\sim$inr/ultracool.html}.

Table \ref{index} lists the values and the derived spectral types measured
from the TiO5, VO-a, and PC3 indices, the spectral type determined by direct
visual comparison, and the spectral type finally adopted for each source.
The adopted spectral type was arrived at as follows. First, the mean value
of the three independent indices was calculated. This value was adopted unless
it was seen to be in gross disagreement with that determined from the template
comparison, in which case the template comparison value was preferred.
Although most spectral type determinations were indeed in good agreement,
discrepancies were seen for a small number of objects, namely
APMPM J2331$-$2750, SSSPM J2322$-$6358, SSSPM J2059$-$8018,
SSSPM J2353$-$4123, SSSPM J0327$-$4236, and SSSPM J2258$-$4639.
For each of these, the spectral type obtained from comparison with
templates was adopted.

%
%

%
%

%
%

%
%
\subsection{Near-infrared classification}
A total of 27 objects were observed spectroscopically in the near-infrared.
We have used a variety of different near-infrared classification schemes which
were designed to provide a best match to the optical classifications for M
and L dwarfs (Reid et al.\ \cite{reid01};
Tokunaga \& Kobayashi \cite{tokunaga99}; Mart\'{\i}n \cite{martin00}).
The reader is referred to those papers for more details on each scheme.

\begin{figure}[!h]
\begin{center}
\includegraphics[width=9.0cm, angle=0]{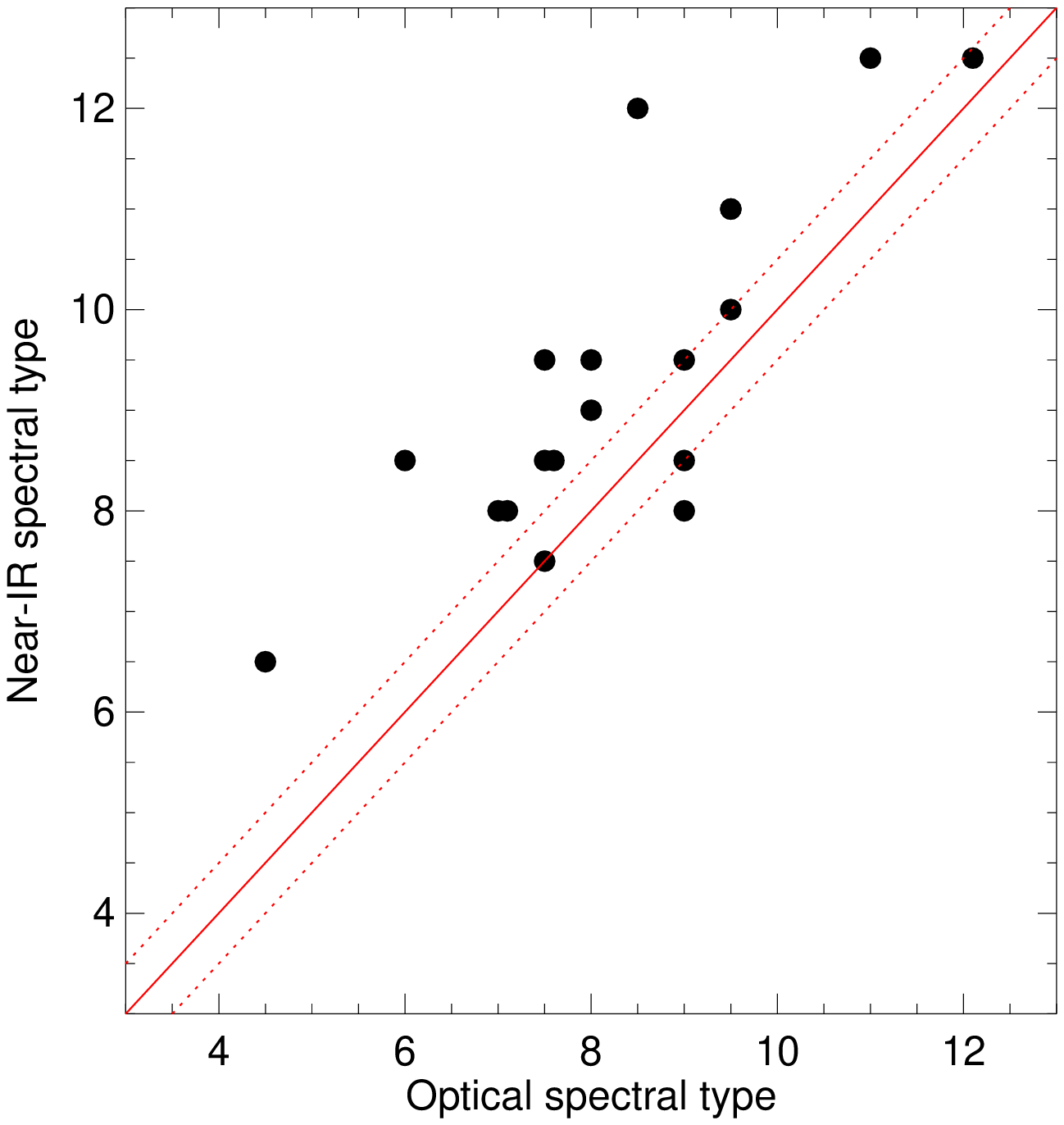}
\caption{
Spectral types determined from our optical spectra vs.\ those obtained from
our near-infrared spectra by taking the mean of the numerical indices of Reid
et al.\ (\cite{reid01}), Tokunaga \& Kobayashi (\cite{tokunaga99}), and
Mart\'{\i}n (\cite{martin00}) (4\,$\equiv$\,M4, \ldots{}, 12\,$\equiv$\,L2).
Some sources have been shifted slightly on the x-axis for clarity.
}
\label{optSpT_NIR}
\end{center}
\end{figure}

Table \ref{NIRindex} lists the numerical indices measured in each scheme,
the corresponding spectral type, and the combined spectral type obtained
by taking their mean for each source. Since 19 of these sources also have
optical spectra, we are able to compare the optical and near-infrared
classification schemes (Fig.\ \ref{optSpT_NIR}). There appears to be a
systematic shift between the schemes, with generally later spectral types
(by roughly one subclass) being assigned in the near-infrared. This 
discrepancy and the relatively large scatter tends to suggest that the
near-infrared schemes are less reliable that their optical counterparts,
somewhat at odds with the findings of other studies 
(see, e.g., Reid et al.\ \cite{reid01}).

%
%

As a consequence, we have adopted spectral types for the sources with
near-infrared spectra in the following manner. For the 19 sources with
optical spectra, we have simply taken the optical classification. For
the remaining 8 sources with near-infrared spectra only, we have ignored
the numerical indices and have instead relied on direct comparison of the
spectra with sources with known spectral types which we observed with the
same setups (SSSPM\,J0306$-$3648, LP\,944-20, Kelu~1), or templates made
available by Neill Reid (see above) and Sandy Leggett\footnote{ 
http://www.jach.hawaii.edu/$\sim$skl/data.html}. The various computed 
indices, the associated spectral types, and the finally adopted spectral
types are listed in Table \ref{NIRindex}. We calculated a formal
uncertainty of half a subclass in the near-infrared classification.
However, due to the systematic observed discrepancy between the optical
and near-infrared spectral types of about one subclass, we adopt
an error of one subclass on the near-infrared spectral types to err
on the cautious side.

%
%
\begin{table*}
\thispagestyle{headings}
\begin{center}
 \caption[NIRindices]{\normalsize
Near-infrared indices and their associated spectral
types derived for the objects observed
spectroscopically either with VLT/ISAAC or NTT/SOFI\@.
The names of the indices refer to the following papers:
Reid et al.\ (\cite{reid01}) (hereafter R01);
Tokunaga \& Kobayashi (\cite{tokunaga99}) (hereafter TK99);
and Mart\'{\i}n (\cite{martin00}) (hereafter M00).
The mean near-infrared spectral type corresponds to the mean
value of the four indices. As discussed in the text and as seen 
in Fig.\ \ref{optSpT_NIR},
the discrepancy between near-infrared and optical classifications
is large. Thus where an optically-derived spectral type exists for a
given source, we have adopted it. When not (sources marked with
note $^{\rm a}$), we have used the spectral type determined by direct comparison
of the near-infrared spectrum with template spectra and/or objects
observed with the same instrument setup, where the spectral types of
the latter were derived from the optical classification schemes.
Preliminary distance estimates based on the near-infrared spectral
types are only given for those objects for which a more accurate
optical spectral type is not yet available.
}
 \scriptsize
 \begin{tabular}{|rl|cc|c|c|c|c|c|c|}
 \hline
RN & Name & \multicolumn{2}{c|}{R01} & TK99 & M00  & Mean  & Adopted & Distance & v$_{t}$ \cr

  & &  H$_{2}$O$^{\rm A}$ & H$_{2}$O$^{\rm B}$ & K1 & Q$_{\rm Hband}$ & NIR SpT  & SpT (optical) & from NIR SpT & \cr
   &      &         &          &    &             &          &         & [pc]         & [km/s] \cr
 \hline
04 & SSSPM J0030$-$3427    &  0.80 (M7.6) &  0.86 (M9.4) & $-$0.06 (M5.9) &  0.55 (M9.0) & M8.0 & M9.0$^{a}$ & 30.4\,$\pm$\,3.6 &  43 \cr
07 & SSSPM J0109$-$5101    &  0.66 (L2.1) &  0.77 (L1.5) &  ~~0.22 (L2.1) &  0.42 (L1.1) & L2.0 & M8.5       &  & \cr
08 & SSSPM J0109$-$4955    &  0.75 (M9.3) &  0.84 (M9.9) &  ~~0.25 (L2.7) &  0.40 (L1.5) & L1.0 & M8.0$^{a}$ & 30.8\,$\pm$\,3.7 &  23 \cr
10 & SSSPM J0124$-$4240    &  0.78 (M8.3) &  0.83 (L0.0) &  ~~0.10 (M9.3) &  0.52 (M9.5) & L2.5 & L0.5$^{a}$ & 17.3\,$\pm$\,2.1 &  22 \cr
11 & SSSPM J0125$-$6546    &  0.76 (M8.9) &  0.94 (M7.2) &  ~~0.02 (M7.6) &  0.63 (M7.6) & M8.0 & M7.0$^{a}$ & 54.0\,$\pm$\,6.5 &  33 \cr
12 & SSSPM J0134$-$6315    &  0.80 (M7.8) &  0.78 (L1.2) &  ~~0.22 (L2.0) &  0.49 (M9.9) & L0.0 & M8.0$^{a}$ & 47.9\,$\pm$\,5.7 &  26 \cr
14 & SSSPM J0204$-$3633    &  0.77 (M8.7) &  1.00 (M5.9) &  ~~0.03 (M7.8) &  0.61 (M8.0) & M7.5 & M7.5       & & \cr
16 & APMPM J0207$-$3722    &  0.80 (M7.8) &  0.83 (L0.0) &  ~~0.02 (M7.7) &  0.52 (M9.5) & M8.5 & M7.0       & & \cr
17 & SSSPM J0215$-$4804    &  0.81 (M7.4) &  0.90 (M8.2) &  ~~0.02 (M7.6) &  0.50 (M9.9) & M8.5 & M8.0$^{a}$ & 30.9\,$\pm$\,3.7 &  52 \cr
18 & SSSPM J0219$-$1939    &  0.60 (M5.7) &  0.69 (L3.4) &  ~~0.16 (L0.7) &  0.41 (L1.3) & L2.5 & L1.0       & & \cr
23 & SSSPM J0306$-$3648    &  0.73 (L0.0) &  0.87 (M9.1) &  ~~0.09 (M9.1) &  0.44 (L0.8) & L0.0 & M8.0       & & \cr
26 & LP 888-18             &  0.74 (M9.7) &  0.86 (M9.3) &  ~~0.08 (M9.0) &  0.45 (L0.6) & M9.5 & M8.0       & & \cr
27 & LP 944-20             &  0.63 (L3.2) &  0.79 (L0.9) &  ~~0.11 (M9.6) &  0.38 (L1.9) & L1.5 & M9.5       & & \cr
28 & LP 775-31             &  0.71 (L0.6) &  0.82 (L0.3) & $-$0.02 (M6.8) &  0.44 (L0.7) & M9.5 & M7.5       & & \cr
29 & LP 655-48             &  0.77 (M8.6) &  0.89 (M8.6) &  ~~0.06 (M8.5) &  0.53 (M9.2) & M8.5 & M7.5       & & \cr
42 & Kelu 1                &  0.66 (L2.2) &  0.75 (L2.1) &  ~~0.28 (L3.2) &  0.35 (L2.3) & L2.5 & L2.0       & & \cr
56 & [HB88] M12            &  0.85 (M6.2) &  1.00 (M5.8) &  ~~0.00 (M7.2) &  0.65 (M7.2) & M6.5 & M4.5       & & \cr
59 & SSSPM J2229$-$6931    &  0.80 (M7.6) &  0.88 (M8.7) & $-$0.14 (M4.2) &  0.55 (M8.9) & M7.5 & L0.0$^{a}$ & 34.4\,$\pm$\,4.1 &  36 \cr
65 & SSSPM J2307$-$5009    &  0.65 (L2.5) &  0.78 (L1.2) &  ~~0.03 (M7.8) &  0.40 (L1.4) & L0.5 & M9.0       & & \cr
66 & SSSPM J2310$-$1759    &  0.72 (L0.3) &  0.78 (L1.4) &  ~~0.13 (L0.0) &  0.37 (L2.0) & L1.0 & M9.5       & & \cr
67 & SSSPM J2319$-$4919    &  0.75 (M9.2) &  0.96 (M6.7) & $-$0.03 (M6.6) &  0.62 (M7.9) & M7.5 & M8.0$^{a}$ & 33.9\,$\pm$\,4.1 &  35 \cr
69 & APMPM J2330$-$4737    &  0.75 (M9.5) &  0.97 (M6.6) &  ~~0.08 (M8.9) &  0.54 (M9.1) & M8.5 & M6.0       & & \cr
70 & APMPM J2331$-$2750    &  0.72 (L0.4) &  0.93 (M7.6) &  ~~0.13 (L0.0) &  0.53 (M9.3) & M9.5 & M7.5       & & \cr
73 & SSSPM J2345$-$6810    &  0.71 (L0.7) &  1.04 (M4.8) &  ~~0.06 (M8.5) &  0.64 (M7.4) & M8.0 & M7.0       & & \cr
75 & SSSPM J2352$-$2538    &  0.70 (L1.0) &  0.92 (M7.9) &  ~~0.10 (M9.3) &  0.48 (L0.1) & M9.5 & M9.0       & & \cr
78 & SSSPM J2356$-$3426    &  0.62 (L3.5) &  0.81 (L0.5) & $-$0.10 (M5.1) &  0.30 (L3.1) & L0.5 & M9.0       & & \cr
79 & SSSPM J2400$-$2008    &  0.63 (L3.1) &  0.82 (L0.4) &  ~~0.07 (M8.7) &  0.34 (L2.5) & L1.0 & M9.5       & & \cr
\hline
 \end{tabular}
 \label{NIRindex}
\end{center}
\end{table*}

%
%
\section{Subdwarfs found in the survey}
\label{subdwarfs}
\begin{figure*}[htbp]
\begin{center}
\includegraphics[width=\linewidth, angle=0]{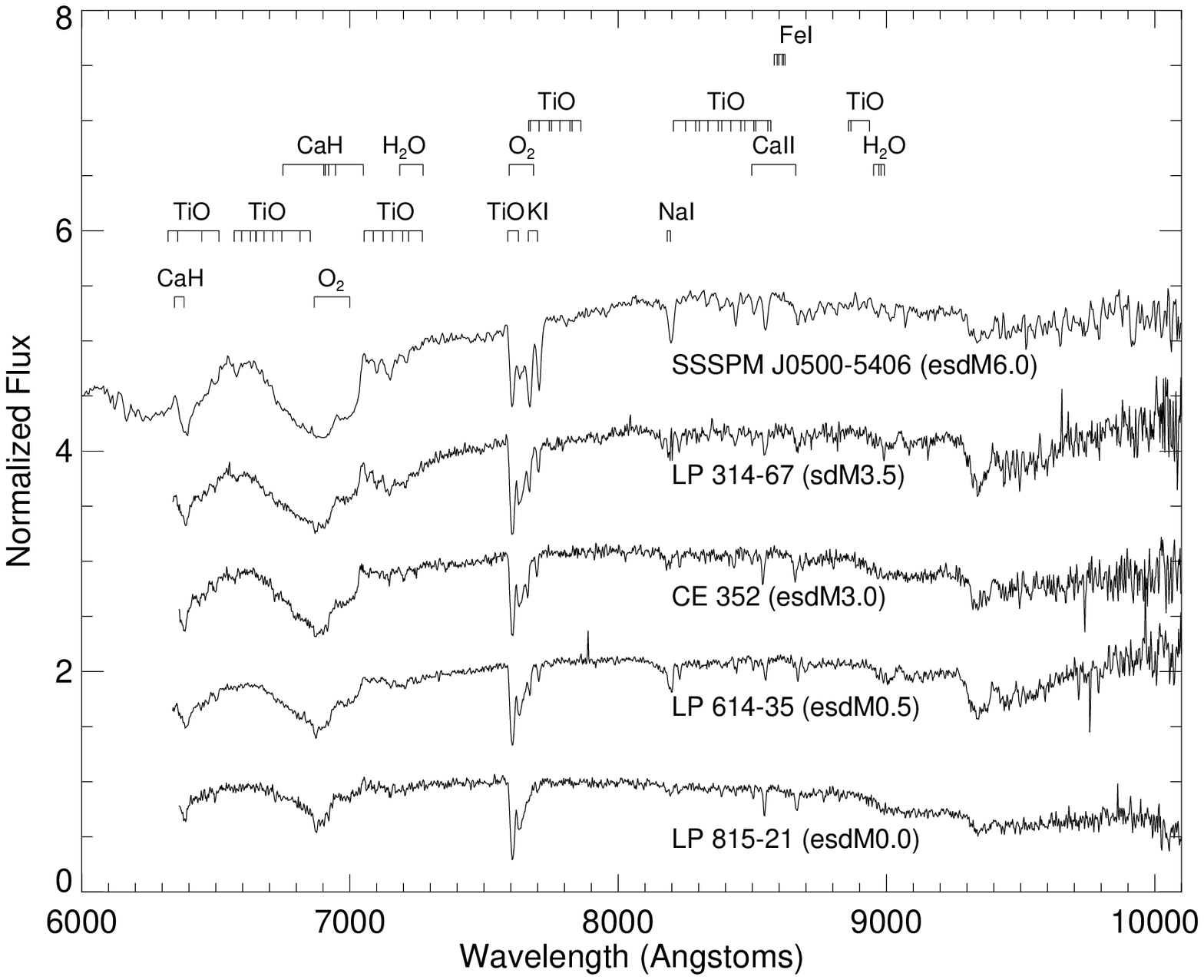}
\caption[Subdwarfs optical spectroscopy]{\normalsize
Optical spectra of the five subdwarfs and extreme subdwarfs found among
the proper motion objects observed with VLT/FORS1 and ESO\,3.6m/EFOSC2.
The classification scheme described in Gizis (\cite{gizis97})
was used to determine the spectral types.
An arbitrary constant has been added to separate the spectra.
}
\label{subdw}
\end{center}
\end{figure*}

We have identified five subdwarfs among our sample of red high proper motion 
objects. The spectral classification scheme for these objects, with lower 
metallicities than those of normal dwarfs, was developed by Reid et al.\ 
(1995) and later built upon by Gizis (\cite{gizis97}). The latter paper
describes the three steps involving TiO and CaH band strengths which are used
to pin down the spectral type. All objects shown in Fig.\ \ref{subdw} fulfil 
the cutoff criteria defined by Eqs.\ 4--6 in Gizis (\cite{gizis97}) which
defines them as subdwarfs and from the indices, three objects (LP\,614-35,
CE\,352, SSSPM\,J0500$-$5406) are unambiguous extreme subdwarfs with spectral 
types esdM0.5, esdM3.0, and esdM6.0, respectively.

SSSPM\,J0500$-$5406 is among the latest extreme subdwarfs known: the current
record holder is the very cool extreme subdwarf, APMPM J0559$-$2903, at
esdM7.0 (Schweitzer et al.\ \cite{schweitzer99}). Very cool normal M-type
subdwarfs (sdM) have been discovered by L\'epine, Shara \& Rich
(\cite{lepine03a}) (LSR\,1425$+$7102; sdM8.0) and by Scholz et al.\
(\cite{scholz04b}) (SSSPM\,J1013$-$1356; sdM9.5), and the first L subdwarfs
have recently been published by Burgasser et al.\ (\cite{burgasser03a}),
L\'epine et al.\ (\cite{lepine03b}), Burgasser et al.\ (\cite{burgasser04b}),
and Scholz et al.\ (\cite{scholz04c}).

%
%

One of the objects found here, LP\,815-21, lies in the boundary region between 
subdwarfs and extreme subdwarfs. A direct comparison with the spectrum of 
LHS\,489 (esdM0.0; Gizis \cite{gizis97}) establishes their similarity and 
allows us to classify LP815-21 also as an esdM0.0 object. Another problematic 
object is LP 314-67, where one of the three indices defined by Gizis
(\cite{gizis97}) locates it between sdM and esdM, one defines it as sdM, and 
the last as esdM: again, direct comparison with a similar spectrum was 
necessary to yield a final spectral type of sdM3.5.

\begin{figure}[h]
\begin{center}
\includegraphics[width=\linewidth, angle=0]{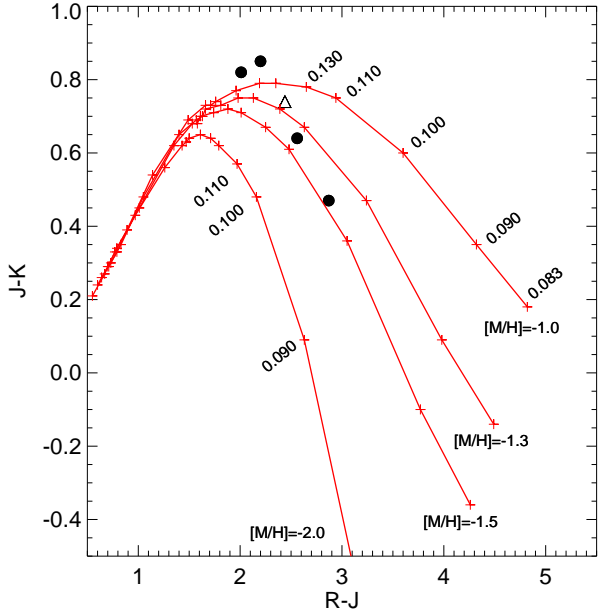}
\caption{
Colour-colour diagram ($R - J$,$J - K$) for the subdwarfs and extreme
subdwarfs found in our proper motion sample. The extreme subdwarfs
are shown as filled circles while the only subdwarf, LP\,314-67, is
displayed as a open triangle. The isochrones from
Baraffe et al.\ (\cite{baraffe97}) for low-metallicity dwarfs 
ranging from [Fe/H]\,=\,$-$2.0 to $-$1.0 are shown for comparison.
Crosses from left to right indicate the following masses:
0.80, 0.75, 0.70, 0.60, 0.50, 0.40, 0.35, 0.30, 0.20, 0.15, 0.13,
0.100, 0.090, 0.085, and 0.083\,M$_{\odot}$. The steps for
the masses are identical for all metallicities.
}
\label{CCD_subdw}
\end{center}
\end{figure}

We have compared the measured colours for each subdwarf with the colours
predicted by the evolutionary models for low-metallicity dwarfs of Baraffe 
et al.\ (\cite{baraffe97}) in order to make a preliminary estimate of
the metallicity and mass for each source, assuming a typical age of 10\,Gyr
(Fig.\ \ref{CCD_subdw}). The derived metallicities tend to be higher
than the assumed metallicity of $-$1.2 and $-$2.0 for subdwarfs and 
extreme subdwarfs proposed by Gizis (\cite{gizis97}). Assuming the 
metallicities from Fig.\ \ref{CCD_subdw}, we have estimated masses of 
0.09--0.10, 0.11, 0.15, 0.10, and 0.20\,M$_{\odot}$ for SSSPM\,J0500$-$5406, 
LP\,314-67, LP\,614-35, CE\,352, and LP\,815-21, respectively.
By comparing the observed magnitudes, metallicities, and masses to
the magnitudes predicted by evolutionary tracks, we have derived distances
of 63\,$\pm$\,9, 143\,$\pm$\,23, 264\,$\pm$\,15, 130\,$\pm$\,23, and
350\,$\pm$\,49\,pc for SSSPM\,J0500$-$5406, LP\,314-67, LP\,614-35, 
CE\,352, and LP\,815-21, respectively, assuming an error of 15\% for these
model-dependent distances.

%

%
\begin{table*}
\thispagestyle{headings}
\begin{center}
 \caption[]{\normalsize
Spectral indices and associated spectral types (with an accuracy of half a subclass),
distance estimates, and kinematics for the five subdwarfs discovered in our survey.
}
 \label{table_subdwarfs}
 \scriptsize
 \begin{tabular}{@{\hspace{1mm}}rl@{\hspace{1mm}}c@{\hspace{3mm}}c@{\hspace{3mm}}c@{\hspace{3mm}}c@{\hspace{3mm}}c@{\hspace{3mm}}c@{\hspace{3mm}}c@{\hspace{3mm}}c@{\hspace{3mm}}c@{\hspace{3mm}}c@{\hspace{3mm}}c@{\hspace{3mm}}c}
 \hline
 \hline
 \smallskip
RN & Name     & TiO5 & CaH1 & CaH2 & CaH3 & SpT & d$_{\rm spec}$ & v$_{t}$  & v$_{r}$  & $U$    & $V$    & $W$   \\
 \smallskip
   &          &      &      &      &      &     & [pc]           & [km/s]    & [km/s]  & [km/s] & [km/s] & [km/s] \\
 \hline
30 & SSSPM J0500$-$5406 & 0.984 & 0.355 & 0.270 & 0.382 &  esdM6.0  & ~~~63\,$\pm$\,09$^a$   &  313    &  $+$247\,$\pm$\,20 & $+$262\,$\pm$\,48  &  $-$287\,$\pm$\,22  &  ~~$-$84\,$\pm$\,17 \\
   &                    &       &       &       &       &           & ~~~70\,$\pm$\,11$^b$   &  348    &                    & $+$294\,$\pm$\,60  &  $-$298\,$\pm$\,25  &  ~~$-$77\,$\pm$\,19 \\
   &                    &       &       &       &       &           & ~~~54\,$\pm$\,08$^c$   &  268    &                    & $+$220\,$\pm$\,43  &  $-$274\,$\pm$\,21  &  ~~$-$94\,$\pm$\,16 \\
35 & LP 314-67          & 0.773 & 0.596 & 0.433 & 0.643 & ~~sdM3.5  &  ~143\,$\pm$\,23$^a$   &  314    &  $+$214\,$\pm$\,20 & $-$128\,$\pm$\,14  &  $-$351\,$\pm$\,52  &  ~~$+$63\,$\pm$\,24 \\
   &                    &       &       &       &       &           &  ~395\,$\pm$\,59$^b$   &  866    &                    & $-$130\,$\pm$\,22  &  $-$872\,$\pm$136   &   $-$115\,$\pm$\,51 \\
   &                    &       &       &       &       &           &  ~193\,$\pm$\,29$^c$   &  423    &                    & $-$128\,$\pm$\,15  &  $-$454\,$\pm$\,71  &  ~~$+$27\,$\pm$\,30 \\
37 & LP 614-35          & 1.057 & 0.744 & 0.660 & 0.806 &  esdM0.5  &  ~264\,$\pm$\,15$^a$   &  237    &  $+$288\,$\pm$\,20 & $-$168\,$\pm$\,35  &  $-$261\,$\pm$\,24  &   $+$206\,$\pm$\,19 \\
   &                    &       &       &       &       &           &  ~265\,$\pm$\,40$^b$   &  237    &                    & $-$169\,$\pm$\,37  &  $-$261\,$\pm$\,25  &   $+$206\,$\pm$\,19 \\
   &                    &       &       &       &       &           &  ~257\,$\pm$\,39$^c$   &  230    &                    & $-$163\,$\pm$\,34  &  $-$258\,$\pm$\,23  &   $+$207\,$\pm$\,19 \\
44 & CE 352             & 0.871 & ---   & 0.471 & 0.626 &  esdM3.0  &  ~130\,$\pm$\,23$^a$   &  217    & ~~$-$66\,$\pm$\,20 & $-$186\,$\pm$\,29  &  $-$121\,$\pm$\,32  &  ~~$-$46\,$\pm$\,11 \\
   &                    &       &       &       &       &           &  ~195\,$\pm$\,29$^b$   &  325    &                    & $-$258\,$\pm$\,65  &  $-$200\,$\pm$\,71  &  ~~$-$52\,$\pm$\,12 \\
   &                    &       &       &       &       &           &  ~149\,$\pm$\,22$^c$   &  249    &                    & $-$207\,$\pm$\,32  &  $-$144\,$\pm$\,34  &  ~~$-$48\,$\pm$\,11 \\
49 & LP 815-21          & 0.952 & ---   & 0.773 & 0.870 &  esdM0.0  &  ~350\,$\pm$\,49$^a$   &  340    &  $+$122\,$\pm$\,20 & $+$269\,$\pm$\,35  &  $-$240\,$\pm$\,53  &  ~~$-$13\,$\pm$\,15 \\
   &                    &       &       &       &       &           &  ~347\,$\pm$\,52$^b$   &  337    &                    & $+$267\,$\pm$\,34  &  $-$237\,$\pm$\,51  &  ~~$-$14\,$\pm$\,15 \\
   &                    &       &       &       &       &           &  ~278\,$\pm$\,42$^c$   &  270    &                    & $+$233\,$\pm$\,30  &  $-$181\,$\pm$\,42  &  ~~$-$23\,$\pm$\,13 \\
 \hline
 \end{tabular}
\end{center}
Notes: $^a$ -- Model-dependent distance estimate according to Baraffe et al.\ (\cite{baraffe97});
       $^b$ -- Distance estimated using comparison objects from Gizis (\cite{gizis97});
       $^c$ -- Distance based on empirical relationship between spectral types and $M_{K_s}$ in L\'epine et al.\ (\cite{lepine03c}).
\end{table*}

As a check, we have also used comparison objects with known spectral types 
and trigonometric parallaxes drawn from Gizis (\cite{gizis97}). By comparing
the 2MASS apparent $JHK_s$ magnitudes for our new subdwarfs with those for
a Gizis (\cite{gizis97}) source with a matching spectral type, we are again 
able to estimate the distances for our sources. We have also used the 
empirical relationships between spectral types and absolute $M_{K_s}$ 
magnitudes for sdM and esdM objects given by L\'epine et al.\ (\cite{lepine03c}).
For the four new extreme subdwarfs,
the bootstrapped distance estimates agree with the 
model-dependent estimates to within the errors, while the bootstrapping 
technique yields significant differences for the subdwarf LP\,314-67, as 
seen in Table \ref{table_subdwarfs}.


Armed with these distance estimates and the proper motions for each source, 
we only require the radial velocity before we are able to calculate the
true space velocities. We have been able to use the relatively sharp Ca\,II 
lines at 8542 and 8662\,\AA{} to detect radial velocity shifts in the five 
subdwarfs. The measured shifts were $+3.5$, $+8.0$, $-2.5$, $+6.5$, and 
$+7.0$\,$\pm$\,0.5\,\AA, corresponding to heliocentric radial velocities 
of $+247$, $+214$, $+288$, $-66$, and $+122\pm 20$\,km/s for
SSSPM\,J0500$-$5406, LP\,314-67, LP\,614-35, CE\,352, and LP\,815-21, 
respectively.


Finally then, we are able to compute the heliocentric space velocities 
following Johnson \& Soderblom (\cite{johnson87}), doing
so for each distance estimation technique as described above. As listed
in Table \ref{table_subdwarfs}, the determined space velocity components 
are typical of (extreme) subdwarfs (cf.\ Figs.\ 17 and 18 in L\'epine et al.\
\cite{lepine03c} with velocities ranging from 100 to 500\,km/s). 
The largest calculated space velocity is that for the 
sdM3.5 object LP 314-67, assuming the distance bootstrapped from the
Gizis (\cite{gizis97}) comparison source: its velocity is comparable to 
the largest space velocities for subdwarfs presented by L\'epine et al.\
(\cite{lepine03c}).

\section{Distance estimates for the normal M and L dwarfs}
\label{dist_est}
Our technique for establishing the distances of the normal M and L dwarfs
in our sample is more straightforward, using the absolute ${\rm M}_J$ versus 
spectral type relationship of Dahn et al.\ (\cite{dahn02}) as follows:
$$ {\rm M}_J = 8.38 + 0.341 \times {\rm SpT},  $$
which is valid for spectral types between M6.5 and L8.0, and where the
spectral type, SpT\,=\,6.5 for M6.5 and increases to 18 for L8. The error
in the relation is $\sim$0.25 magnitudes and in computing the distances for
the sources in our sample, we have combined that error with a 0.1 magnitude
error in the measured apparent $J$ magnitude. Note that we have used
the $J$ magnitudes from 2MASS and not our own ISAAC $J_{s}$ magnitudes
to compute distances.

For objects with spectral types earlier than M6.5, we have employed the 
absolute magnitudes given by Kirkpatrick \& McCarthy (\cite{kirkpatrick94}) 
for all early- and mid-M dwarfs, except for M5.5 dwarfs. For these, we used
absolute magnitudes of M$_{J}$\,=\,9.87, M$_{H}$\,=\,9.31, and 
M$_{Ks}$\,=\,8.97 as kindly provided by Hartmut Jahrei\ss{} (personal
communication), based on the following standards: LHS\,2, LHS\,39, LHS\,549, 
LHS\,1565, and LHS\,3339. For the early type sources in our sample, the
distances were determined as an average of those calculated for each of the
three near-infrared filters, and the uncertainties were estimated using the 
assumptions described above for the late-type objects.

The resulting distance estimates are listed in Tables \ref{index} and 
\ref{NIRindex}. The great majority of the late-type ($>$\,M6) dwarfs 
appear to be located within 50\,pc of the Sun, with 24 falling in the
range of the catalogue of nearby stars ($<$\,25\,pc) and 11 of those
within 15\,pc. On the other hand, the majority of the earlier-type objects 
are located at larger distances ranging from 100--200\,pc: only three of 
them are members of the 25\,pc sample.

\section{Kinematics and activity}
\label{kinact}
\begin{figure}[htb]
\begin{center}
\includegraphics[width=6.0cm, angle=270]{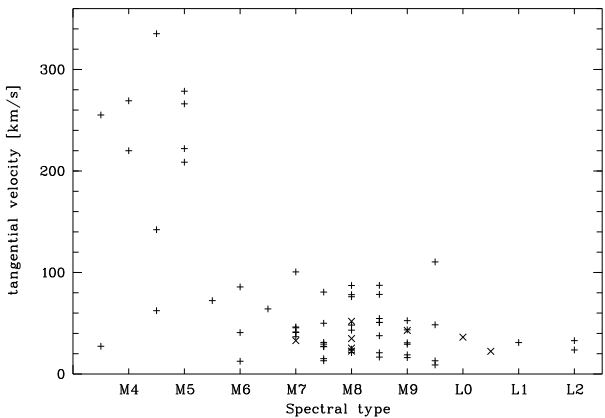}
\caption[Tangential velocity vs. Spectral Type]{ 
Tangential velocities vs.\ spectral types for all spectroscopically
classified ($+$ optical, $\times$ near-infrared) M and L dwarfs in this study.
}
\label{vtSpT}
\end{center}
\end{figure}

Combining the distance estimates with the measured proper motions, the
tangential velocities of the sources can be calculated (Tables \ref{index}
and \ref{NIRindex}). The large distances estimated for most of the early-type 
M dwarfs ($<$\,M5.5) and the relatively large proper motions measured
in some cases lead to extremely large tangential velocities of 
$\sim$\,200--300\,km/s (Fig.\ \ref{vtSpT}), comparable with those of the 
subdwarfs described earlier (Table \ref{table_subdwarfs}). The most extreme 
example is the M4.5 dwarf, APMPM\,J2036$-$4936, with a tangential velocity of 
335\,km/s. Such large velocities are more typical of 
metal-poor stars than of metal-rich ones,
although L\'epine et al.\ (\cite{lepine03c}) have also reported several
normal M dwarfs which may be members of the Galactic halo population.
Gizis (1997) described a number of sources with early M-type
spectra and halo kinematics which he classified as normal M dwarfs,
rather than sdM or esdM subdwarfs. Indeed such objects are to be expected,
since his classification scheme sources with metallicities
of $\approx -0.5$ are normal M dwarfs, not subdwarfs, and the
metallicity distribution of halo stars extends down to this level.
Among the later-type ($>$\,M5.5) dwarfs, the tangential velocities are
typically substantially less than 100\,km/s, with the M9.5 dwarf, 
SSSPM\,J2400-2008, the most extreme case at 110\,km/s.

Gizis et al.\ (\cite{gizis00}) have demonstrated a relationship 
between tangential velocity and chromospheric activity for late (M8--M9.5)
dwarfs, inasmuch as all objects with H$\alpha$ equivalent widths over 10\AA{}
have tangential velocities greater than 20\,km/s and up to 80\,km/s. The 
velocity dispersion of a population of stars increases with age as they 
interact with the potential of the galactic disk and thus low velocity stars 
are generally younger than high velocity stars. The usual prejudice is that 
young, more rapidly rotating stars exhibit more chromospheric activity than 
their older, spun-down counterparts, but the Gizis et al.\ (\cite{gizis00}) 
finding shows that in fact the opposite applies to ultracool dwarfs, where
the activity declines due to the very low temperatures. 

There appears to be a second order effect however, namely that activity is 
also correlated with mass. For a given ultracool spectral type (e.g., late M 
or L), a sample of sources may contain both old and young objects: the old 
objects will be true stellar sources (i.e., above $\sim$0.075\,M$_\odot$) 
which have cooled and contracted down to their main sequence parameters, 
while the younger sources could well be substellar objects (i.e., below 
$\sim$0.075\,M$_\odot$) which are still deriving luminosity and an enhanced 
effective temperature from their pre-main sequence contraction. These latter
sources will be slightly larger and more luminous, and have a lower surface 
gravity than their older, same-spectral-type counterparts. Gizis et al.\
(\cite{gizis00}) suggest that this may lead to the observed reduction in
chromospheric activity.

With our new sample of ultracool dwarfs, we can explore these relationships
a little further, although it must be emphasised that we cannot do so in a
statistical sense. Our sample is proper-motion selected and is therefore
biased towards older, high velocity objects and away from younger, lower
velocity sources.

We have used two different indices to estimate the level of chromospheric
activity for the objects in our sample. First, we have computed the H$\alpha$ 
index defined by Kirkpatrick et al.\ (\cite{kirkpatrick99}; see their Table 4),
which measures the strength of the H$\alpha$ emission line compared to the 
neighbouring continuum: values above 1 indicate significant emission. Second, 
we have measured the equivalent width of the H$\alpha$ line. The results of 
both measurements are given in Table~\ref{index}. 

%
%
\begin{figure}[htb]
\begin{center} 
\includegraphics[width=6.0cm, angle=270]{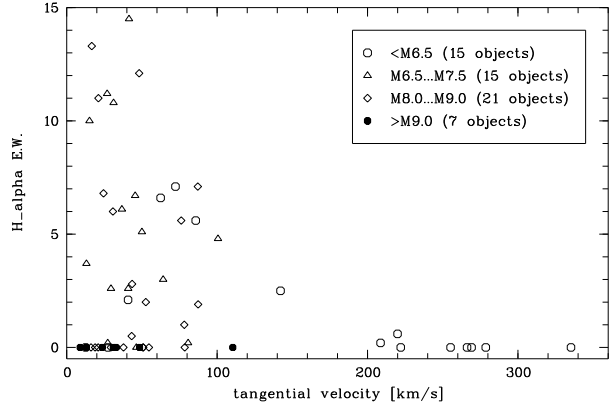}
\caption[Halpha EW vs. Tangential velocity]{
H$\alpha$ equivalent widths vs.\ tangential velocities for all 
M dwarfs in the present study.
}
\label{HaEWvt} 
\end{center}
\end{figure}

First, we note that there is a peak in activity around spectral types M6--M7,
as also reported by Hawley, Gizis, \& Reid (\cite{hawley96}) and Gizis et al.\
(\cite{gizis00}) for larger samples of nearby M dwarfs. In more detail, we see
that there are seven very active objects with H${\alpha}$ equivalent widths
larger than 10\,\AA{}, all of which are M7--M8.5 dwarfs. Interestingly, these
sources all have relatively small tangential velocities, $v_t < 50$\,km/s,
three of them below 20\,km/s: their average velocity is roughly half that of
the full population of M7--M8.5 sources in our sample.

Thus, at first sight, this subsample appears to contradict the finding of
Gizis et al.\ (\cite{gizis00}), namely that active, very late type sources
tended to have larger velocities than their inactive counterparts. However,
when we more rigorously compare like with like, we find that their conclusion
remains valid. If we consider only the M8--M9.5 sources and divide them 
into H$\alpha$ active (EW$\geq$0.5\AA; 12 sources) and inactive
(EW$<$0.5\AA; 13 sources) samples, we find mean tangential velocities
of 52 and 40.5\,km/s, respectively. That is, the active sources have
higher velocities on average, consistent with the finding of 
Gizis et al.\ (\cite{gizis00}).

%
%
\section{Discussion related to specific objects}
\label{special_obj}
\subsection{L dwarfs}
\label{Ldwarfs}


In Lodieu et al.\ (\cite{lodieu02}), we described the discovery of three new 
early L dwarfs from our proper motion survey, with optical and near-infrared
spectroscopy for two of them (SSSPM\,J2310$-$1759, SSSPM\,J0219$-$1939) and 
near-infrared spectroscopy alone for the third (SSSPM\,J0109$-$5101). The
spectral types derived by Lodieu et al.\ (\cite{lodieu02}) based on 
near-infrared spectral indices were L1, L2, and L2, respectively, but we 
now find that these results must be revised.

We have made direct comparisons of the optical spectra which we now have for 
all three sources with template stars observed with the same instrumental 
set-up, deriving revised spectral types of M9.5, L1.0, and M8.5 for
SSSPM\,J2310$-$1759, SSSPM\,J0219$-$1939, and SSSPM\,J0109$-$5101, respectively, 
with uncertainties of half a subclass. These differ significantly from the
previous results, in line with the trend shown in Fig.\ \ref{optSpT_NIR}, 
where the near-infrared spectral classification tends to yield a later 
spectral types than found from the optical spectra.

As a consequence of the revised spectral types, the corresponding spectroscopic
parallaxes shown in Table \ref{index} have also been changed, although the 
objects remain relatively nearby: SSSPM\,J2310$-$1759 and SSSPM\,J0219$-$1939
lie within 30\,pc, while SSSPM\,J0109$-$5101 may be within 15\,pc.

The nearest L dwarf in our sample (SSSPM J0829$-$1309; L2.0) has a 
spectroscopic distance of $\sim$\,12\,pc and we confirm the spectral type 
obtained from a lower signal-to-noise spectrum in the discovery paper by 
Scholz \& Meusinger (\cite{scholz02b}). Its optical spectrum is shown in 
Fig.\ \ref{latest} alongside the very similar spectrum for Kelu 1.

Finally, there are two other L dwarf candidates (SSSPM\,J0124$-$4240 at L0.5; 
SSSPM\,J2229$-$6931 at L0.0) in our sample for which, however, only 
near-infrared spectra could be used for classification (see Fig.\ \ref{specIR}).
Given the apparent systematic shift in spectral types between optical and
near-infrared spectra for the sources discussed above and as similarly
seen for Kelu 1, LP\,944-20, and SSSPM\,J2400$-$2008, these early L dwarf 
classifications for SSSPM\,J0124$-$4240 and SSSPM\,J2229$-$6931 must be 
regarded with some caution: it is entirely possible that they will be
reclassified as late M dwarfs once appropriate optical spectra are obtained.

%
%
\subsection{Nearby M dwarfs}
\label{Mdwarfs}
Our sample contains two late M dwarfs within 10\,pc, LP\,775-31 and LP\,655-48, 
which we originally classifed as M8.0 and M7.5, respectively, on their
discovery (McCaughrean et al.\ \cite{mjm02}). Cruz \& Reid (\cite{cruz02})
had earlier claimed that both sources were M6 dwarfs, but Cruz et al.\ 
(\cite{cruz03}) subsequently reclassified both as M7. We have also revisited 
their optical spectra as part of the present work and adopt spectral types of 
M7.5 for both (Table \ref{index}), consistent with the Cruz et al.\ 
(\cite{cruz03}) results to within the errors. In either case, the distances
nominally stay below 10\,pc, but as pointed out by Cruz et al.\ 
(\cite{cruz03}), trigonometric parallaxes are needed to settle the matter.

However, it is worth noting that the near-infrared spectra for these two
sources differ significantly, with LP\,655-48 being redder in the range
1.46--1.69\,$\mu$m, redder from 2.0--2.2\,$\mu$m, bluer from 2.2--2.4\,$\mu$m,
and exhibiting a stronger CO bandhead break. As recently demonstrated by
Close et al.\ (\cite{close03}), these differences may indicate that one or
both of the objects may be a binary with a very late type companion.

In addition to these very nearby sources, there are several other M dwarfs
in our sample which we find to be rather nearby, including LP\,888-18 and 
APMPM\,J1251$-$2121 at $\sim$\,11\,pc, and SSSPM\,J0829$-$1309 at 
$\sim$\,12\,pc.
 
%
%
\subsection{Objects with proper motions larger than 1 arcsec per year}
\label{1asyr}
Four of the discoveries presented in this study have proper motions exceeding 
1 arcsec/yr. One of these objects, SSSPM\,J0500$-$5406, was spectroscopically 
classified as an ultracool subdwarf (esdM6.0), while the other three are
mid-M dwarfs: SSSPM\,J1926$-$4311 (M5.0), APMPM\,J1957$-$4216 (M5.0), and 
APMPM\,J2330$-$4737 (M6.0). The large tangential velocities ($>$\,200\,km/s) 
implied for SSSPM\,J1926$-$4311 and APMPM\,J1957$-$4216 suggest that they
may well be members of the Galactic halo population, while the derived
$v_t$\,$\sim$\,90\,km/s for APMPM\,J2330$-$4737 may make it a representative
of the thick disk. The identification of a relatively large number of 
these relatively rare objects in this study simply confirms the strong bias 
towards them that naturally results from proper motion based searches (see 
also Scholz et al.\ \cite{scholz04b}). 

%
\subsection{Objects with blue colours}
\label{coloutl}
There are two objects in our sample with unusually blue colours for their
spectral types, even after considering the rather large dispersion in the
colours originating with the relatively poor accuracy of the photographic
SSS magnitudes (see Fig.\ \ref{colSpT}). The $B_J - R$ colour of the first
of these objects, SSSPM\,J0027$-$5402, is $\sim$\,1.1, whereas its M7
spectral type would predict something closer to 2.5. None of the other 
colours for this source seem out of character with its spectral type which 
would, at first sight, suggest a photometric error in $B_J$, despite its
apparent brightness at that wavelength. An alternative solution might be
the presence of a very blue companion or chance aligned field star.

The second source, SSSPM J0134$-$6315, has been classified as M8, but its
$I - J$\,$\sim$\,1.5 and $R - K_{s}$\,$\sim$\,5.3 are more consistent with 
something closer to M5. In this case, the explanation may lie with the fact
that we have only a near-infrared spectrum for this source, which may again
have skewed our classification towards a later type than would be derived
in the optical, as discussed above.

There are two other objects labelled in Fig.\ \ref{colSpT} with nominally
erroneous colours: these are not members of our sample, but are shown for
comparison. One is 2MASS\,1411$-$21, classified as an M9 dwarf by both Cruz 
et al.\ (\cite{cruz03}) and Kendall et al.\ (\cite{kendall04}), but also
appearing very blue in the optical with $B_{J} - R$\,$\sim$\,2.0. Again, a 
blue companion or aligned field star might be the explanation. The other 
source, SDSS\,1326-00 (Fan et al.\ \cite{fan00}) has 
an uncertain spectral type of ``L8?''
but was barely detected in the SSS $I$ band data and has a very small $I - J$
colour of 2.2, much below the expected $\sim$\,3.5. The proper motion for
this source of ($-$294,$-$1027)\,$\pm$\,(20,3)\,mas/yr obtained from one SSS, 
one 2MASS, and one SDSS position formally supports its identification, but
there remains room for error, as the epochs of the three observations 
were not well distributed (SSS 1991.256; 2MASS 1999.093; SDSS 1999.219). 
This object requires further investigation and at the very least, a new 
epoch position would prove very helpful in verifying the proper motion and 
the corresponding SSS identification.

%
\subsection{Common proper motion stars}
\label{newcpm}
One of the new high proper motion objects was found to be a common proper 
motion companion 32 arcsec from the previously known LHS\,3141 and has 
therefore been named LHS\,3141B\@. The catalogued proper motion of the primary,
($+$204,$-$519)\,$\pm$\,(5,8)\,mas/yr, agrees to within the errors with our
measured proper motion for LHS\,3141B\@. We have classified this latter 
object as an M7 dwarf, but unfortunately, no spectral type presently exists 
for the brighter primary. According to its optical-to-infrared colours
($I - J$\,$\sim$\,1.1; $R - K_{s}$\,$\sim$\,3.4), it should be an early-type 
M dwarf. Assuming a distance of 37\,pc as derived from the spectroscopic
parallax for the secondary, the projected physical separation is 
$\sim$\,1200\,AU\@.

The previously known common proper motion pair LDS\,4980 was included in our
spectroscopic program due to the relatively red colour measured from APM 
scans of the POSS-1 $O$ and $E$ plates. The spectral types determined for
the pair, M3.5 and M4, are rather early and imply a relatively large distance
of $\sim$\,210\,pc, in good agreement with independent estimates for the two
components. The measured angular separation of about 79 arcsec between the
pair then translates into a large projected physical separation of 
$\sim$\,16\,600\,AU or 0.08\,pc. The proper motions of the two components
also agree within the errors and the implied tangential velocity appears
to be in excess of 200\,km/s, rather large for M dwarfs and indicating 
membership in the Galactic halo.

Finally, our sample contains another new common proper motion component to a 
known system, APMPM\,J2354$-$3316, recently been described in detail by 
Scholz et al.\ (\cite{scholz04a}). It is a wide M8.5 companion to the M4/DA 
binary LHS\,4039/LHS\,4040 and was found to show very strong H$\alpha$ emission 
line and a blue continuum below 7500\,\AA.

%
\subsection{Objects without spectroscopy}
\label{nospec}
As discussed in \S\ref{obs}, eight of the objects in our sample have not
yet been observed spectroscopically at either optical or near-infrared
wavelengths, although imaging observations were made for all of them using 
the VLT\@. 

Five of these objects (APMPM\,J0207$-$7214, APMPM\,J0244$-$5203, 
LHS\,2555a, APMPM\,J0331$-$2349, APMPM\,J1212$-$2126) have small 
optical-to-infrared indices ($R - K_{s}$\,$<$\,4 and $I - J$\,$<$\,1.5) 
typical of early-M dwarfs (cf.\ Fig.\ \ref{colSpT}) or subdwarfs 
(\S\ref{subdwarfs}). The other three objects (APMPM\,J0057$-$7604,
APMPM\,J0232$-$4437, APMPM\,J0536$-$5358) have larger indices
(5.2\,$<$\,$R - K_{s}$\,$<$\,6.2 and $I - J$\,$\sim$\,2.1) similar to the 
spectroscopically-classified M5--M7 dwarfs in our sample. The brightest of 
these objects, APMPM\,J0057$-$7604 with $K_{s}$\,$\sim$\,11.6, is most likely 
located within 50\,pc and could be within 25\,pc sample if its spectral type 
turns out to be later than M6.

%
\section{Summary and outlook}
\label{summary}
We have presented optical and near-infrared photometry and spectroscopy for 
a sample of 71 red proper motion objects and we have classified 58 and 27
of them using optical and near-infrared spectroscopy, respectively; the
remaining eight sources have optical and near-infrared photometry only.

The spectroscopic sample comprises 60 M dwarfs, 6 L dwarfs, and 5 subdwarfs. 
For the sources with both optical and near-infrared spectroscopy, we have
observed a systematic shift in the near-infrared of about one subtype 
towards later spectral types. As a consequence, we have favoured the optical
spectral types whenever possible for consistency with other approaches in the
literature. Partly in connection with this, we have revisited the spectral 
typing of three previously published L dwarfs and two M dwarfs in the course 
of this work, generally finding them to have earlier types than before. 

The large majority of the late-type dwarfs (spectral types later than M6) are
located within 50\,pc of the Sun and generally have small tangential 
velocities ($<$\,50\,km/s), whereas the early-type M dwarfs tend to have 
larger distances ranging from 100--200\,pc and tangential velocities in the 
range 100--200\,km/s. These correlations, of course, simply reflect the search 
strategy which we have employed, which is flux- and proper motion-limited. 
Intrinsically fainter sources can be nearby and need only have relatively
low velocities; brighter, earlier-type stars can be picked up at larger
distances, as long as they have larger space velocities. 
We see the previously known peak in chromospheric activity around M6--M7 
and also provide mild confirmation for the finding by Gizis et al.\ (\cite{gizis00})
of a correlation between tangential velocity and chromospheric activity over
the restricted spectral type range M8--M9.5\@. However, as discussed
in \S\ref{kinact}, our proper motion selection strategy is biased towards
older, higher velocity sources, making it hard to assign any kind of
statistical significance to the reality of this effect.

Among the subdwarfs in the sample, four are classified as extreme subdwarfs 
and one of these, SSSPM\,0500$-$5406 is among the coolest M subdwarfs 
discovered to date with an estimated spectral type of esdM6. The fact that 
we have not discovered any new L subdwarfs perhaps suggests that Population\,II 
may be deficient in substellar objects with respect to Population\,I, i.e.,
the local field. This is contrary to expectations: very metal-poor gas has 
a reduced dust cooling efficiency (less effective gas-to-dust coupling)
and thus should yield smaller minimum Jeans mass fragments. These should then
be able to collapse to form brown dwarfs before they merge into larger 
objects, and thus in theory yield large numbers of substellar objects 
(Zinnecker \cite{zinnecker95}). However, the search strategies for
Population\,II brown dwarfs (substellar subdwarfs) still suffer from
various selection biases, with the  general problem of identifying
very large proper motions at the magnitude limit of optical surveys
and difficulties in optimising the colour search parameters.

Finally, we have identified several interesting new sources worthy of more
detailed follow-up observations, including trigonometric parallax 
measurements, spectroscopy in the lithium line to search for evidence of
a substellar nature, searches for low-mass companions, and so on. Three of the
new objects lie at close to 10\,pc; four objects have proper motions larger 
than 1 arcsec/yr; two targets have been identified as common proper motion
companions to previously known nearby objects; and some objects exhibit 
unusual blue colours, which might indicate the presence of a companion.

\begin{acknowledgements}
Our survey for new, faint, high proper motion objects is based on the 
SuperCOSMOS Sky Surveys, i.e., digitised data obtained from scans of UKST 
and ESO Schmidt plates. We would like to thank the SuperCOSMOS team for 
producing such excellent data. All of the new observations presented in 
this paper were obtained with ESO's telescopes on Paranal and La Silla,
and we acknowledge the expertise of, advice from, and enriching discussions 
with the ESO support teams during the various runs. We thank the referee, 
John Gizis, for his kind report and some useful comments. This research
has also made use of data products from the Two Micron All Sky Survey, which 
is a joint project of the University of Massachusetts and the Infrared 
Processing and Analysis Center, funded by the National Aeronautics and Space 
Administration and the National Science Foundation, as well as of the VizieR 
catalogue access tool, CDS, Strasbourg. We thank Gyula Szokoly for obtaining
low-resolution observations of the objects APMPM\,J2036$-$4936 and LP\,859-1 
for us. Finally, NL and MJM thank the European Research Training Network 
on ``The Formation and Evolution of Young Stellar Clusters'' 
(HPRN-CT-2000-00155) for financial support.
\end{acknowledgements}

\end{document}